\documentclass[numberedappendix,twocolumn]{openjournal}

\shorttitle{Curious case of Centaurus A}
\shortauthors{Weerasooriya et al.}
\graphicspath{{./}{figures/}}
\usepackage{lastpage}
\usepackage{comment}
\usepackage[caption=false]{subfig}
\usepackage{amsmath}
\usepackage{xcolor}
\usepackage{orcidlink}
\usepackage{hyperref}
\begin{document}

%\title{Simulating Dandelions of the Milky Way: Using \textsc{Galacticus} to Model Dwarf Satellites of the Milky Way}

\title{The Curious Case of Centaurus A: On the Subject of the Quenched satellites}

\author{Sachi Weerasooriya\orcidlink{0000-0001-9485-6536}$^*$}
\affiliation{The Observatories of the Carnegie Institution of Science, 813 Santa Barbara St, Pasadena, CA 91101, USA}
\altaffiliation{$^*$sweerasooriya@carnegiescience.edu}
% \affiliation{Department of Physics and Astronomy, Texas Christian University, Fort Worth, TX 76109, USA}
\author{Mia Sauda Bovill}
\affiliation{Department of Astronomy, University of Maryland, College Park, MD 20742}

\author{Matthew A. Taylor\orcidlink{0000-0003-3009-4928}}
\affiliation{ University of Calgary,
2500 University Drive NW,
Calgary Alberta T2N 1N4,
CANADA}

 \author{Andrew J. Benson\orcidlink{0000-0001-5501-6008}}
\affiliation{Carnegie Science Observatories, 813 Santa Barbara St, Pasadena, CA 91101, USA}

\author{Cameron Leahy\orcidlink{0000-0002-0091-9610}}
\affiliation{ University of Calgary,
2500 University Drive NW,
Calgary Alberta T2N 1N4,
CANADA}

\author{Alexis Vazquez}
\affiliation{San Francisco State University, CA 94132, USA}

% \author{Cameron Leahy}
\author{Niusha Ahvazi\orcidlink{0009-0002-1233-2013}}
\affiliation{University of Virginia, 530 McCormick Road, P.O. Box 400325, Charlottesville, VA 22904, USA}
 \affiliation{Carnegie Science Observatories, 813 Santa Barbara Street, Pasadena, California 91101, USA}
  
 \author{Pamela M. Marcum}
\affiliation{NASA Ames Research Center, Moffett Field, CA 94035, USA}
\author{Alejandro S. Borlaff\orcidlink{0000-0003-3249-4431}}
\affiliation{Bay Area Environmental Research Institute, Moffett Field, California 94035, USA}
\affiliation{NASA Ames Research Center, Moffett Field, CA 94035, USA}

\begin{abstract}
 
The satellite system of Centaurus A presents a curious cosmological puzzle: while the global population is consistent with theoretical expectations, its inner regions ($d<150~\mathrm{kpc}$) exhibit a deficit of luminous satellite galaxies. Using the \textsc{Galacticus} semi-analytic model (SAM) applied to high-resolution N-body merger trees, we investigate potential quenching mechanisms to explain this trend.  Our fiducial models, calibrated to the Milky Way, reproduce the overall Cen A population but overpredict the number of bright inner-halo satellites by a factor of $4\pm1$ at $\mathrm{M_V}<-15.8$. We find that this is not due to statistical variance. Instead, the spatial coincidence of this deficiency with Cen A's massive, kiloparsec-scale radio lobes suggests a powerful environmental driver. We explore a range of physical scenarios, including enhanced tidal disruption, reionization quenching, and suppressed accretion into halos from the surrounding intergalactic medium. Our results indicate that AGN-driven thermal feedback at $z < 5$ can significantly suppress star formation in satellites, effectively truncating the bright end of the inner luminosity function. Our work suggests that the ``Curious Case of Centaurus A" may provide evidence of AGN feedback within the host galaxy that regulates the survival and evolution of its dwarf galaxy satellites.

% evidence of AGN feedback of external AGN of the host that influences their survival and evolution of its surrounding dwarf galaxies.
    
\end{abstract}
\section{Introduction}\label{sec:intro}

Due to their very shallow potential wells, dwarf galaxies are excellent probes of both internal and environmental feedback and quenching processes \citep[{\it e.g.},][]{dekel1986,thoul1996,benson2002a,okamoto2010,Sawala+2016,Sales+2022,Kim+2025}. Such quenching mechanisms include tidal stripping, ram pressure stripping \citep{Simpson+2018}, and stellar and Active Galactic Nuclei (AGN) feedback from the host, among others. To date, few studies have explored the impact of AGN in the central galaxy on its satellites \citep{Dashyan+2019,Das&Pandey2025,Visser-Zadvornyi+2025}.
    
 Until recently, detailed observational \citep{Richardson+2011,mcconnachie2012,Drlica-Wagner+2020,Geha+2024} and theoretical \citep{Cole+2014,Applebaum2021,akins2021,Agertz+2020,Nadler+2024,Bose&Deason2025}  studies of dwarfs have been limited to the Local Group (LG) satellites and isolated dwarfs \citep{Kim+2024}. Forthcoming telescopes, including the Nancy Grace Roman Space Telescope, the Euclid space telescope \citep{Euclid2025}, and the Vera C. Rubin observatory \citep{LSST2025}, will soon move us into an era in which we can study dwarf galaxies in an unprecedented range of environments, expanding upon existing studies of dwarfs in the Local Volume \citep{Ferrarese+2012,Danieli+2018,Venhola+2019,Liu+2020,Danieli+2023,Christensen+2024}. In these AGN rich denser environments, high-temperature circumgalactic media may strongly influence star formation, a process that is still poorly understood. \\
 In advance of these new data, progress can be made by exploring other environments accessible to present observations. Centaurus A (Cen A) is the closest easily observable giant elliptical galaxy, located 3.8 Mpc from the Milky Way \citep{Harris+2010}. Although its total halo mass is somewhat poorly constrained, with estimates from $\sim4.7\times10^{12}\,\mathrm{M}{\odot}$ \citep{Pearson+2022}, $\sim7.3\pm2\times10^{12}\,\mathrm{M}{\odot}$\citep{Faucher+2025} to $1.8\times10^{13}\,\mathrm{M}{\odot}$ \citep{van2000}, Cen A clearly lies between the Local Group and cluster scales. Extensive investigations have been conducted on the AGN of Cen A for many years \citep{Meier+1989, Tingay+1994, Jones+1996, Kellermann+1997, Tingay+1998, Fujisawa+2000, Tingay&Murphy2001, Tingay+2001, Horiuchi+2006, Muller+2011, Muller+2014,Borkar+2021}, finding that Cen A is home to a Fanaroff-Riley Type I (FR-I) source with a total power $P_{2.7\,\mathrm{GHz}}=1024.26\,\mathrm{W\,\mathrm{H{z}}^{-1}}$ \citep{Struv+2010b}. The radio jet extends from sub-parsec scales \citep{Jones+1996, Horiuchi+2006} to 6~kpc from the nucleus. The Cen A supermassive black hole (SMBH) is actively accreting material and producing a jet that is detected in X-ray emission \citep{Kraft+2002} and radio continuum \citep{Schreier+1979,Schreier+1981}. The radio structure of Cen A consists of asymmetric inner lobes and asymmetric structures toward the outer parts. The middle lobe extends to 40~kpc, while the outer lobe extends to more than 500~kpc \citep{Morganti+1999}. The northern radio lobe of Cen A extends for $\sim300\,\mathrm{kpc}$ in the north-south direction and $\sim 200\,\mathrm{kpc}$ in the east-west direction \citep{Stawarz+2013}. The above studies indicate that the lobes extend well into and perhaps beyond the halo, suggesting that the AGN could plausibly influence satellite galaxies of Cen A both before and after they fall into its halo. \\
 Multiple recent surveys of Cen A \citep[NGC 5128,][]{crnojevic2010,crnojevic+2011,crnojevic2014,crnojevic2016,crnojevic2019,SCABS,Taylor+2018,muller2015,muller2017,Muller2019} have discovered dwarf galaxies in the central regions of the halo. These observations complement the next generation telescopes making Cen A an ideal target for detailed study. Moving from the mass scale of the LG to that of Cen A, AGN begin to play a major role in governing feedback within massive ($\geq\,10^{12}\,\mathrm{M}_{\odot}$) galaxies \citep{Bazzi+2025,Bugiani+2025}. The AGN hosted by Cen A was potentially activated by a major merger event that occurred $\sim 2$ Gyr ago \citep{Wang+2020}, which, in addition to the proximity of the system, provides a unique opportunity to study the connection between an AGN and its surrounding satellite population.

Cen A analogs (which we define here as systems that have comparable halo mass to the observed Cen A, and without any halos of mass greater than $M_\mathrm{halo}=10^{12}\,\mathrm{\mathrm{M}_{\odot}}$ within 1.4 Mpc) exist in hydrodynamic simulations such as TNG100 and those of \cite{Wang+2020}, who reproduced some sub-structures of Cen A and the major mergers that occurred a few Gyr ago. \cite{Muller2019} have analyzed Cen A analogs in TNG100, which includes AGN physics, to study Cen A satellites, identifying a lack of bright satellites and an overall overestimation of satellites within $150\,\mathrm{kpc}$ in simulations compared to observations. However, \cite{Muller2019} determined that the Cen A satellite luminosity function falls within the 90\% interval of halo-to-halo variations among Cen A analogs in the TNG100 simulation (across all $M_\mathrm{V}$), and that the over-prediction of fainter satellites may be caused by a lack of convergence due to the limited resolution in modeling fainter satellites in TNG100. They further compare satellites within $150\,\mathrm{kpc}$ for Cen A analogs of differing masses: $5\times10^{12}\,\mathrm{M}_{\odot}$, $7\times10^{12}\,\mathrm{M}_{\odot}$, $9\times10^{12}\,\mathrm{M}_{\odot}$, and $11\times10^{12}\,\mathrm{M}_{\odot}$). The authors find that the number of satellites per absolute V band magnitude increases as the mass of the Cen A analog increases, but find no change in the slope of the luminosity function.
    
Although hydrodynamic simulations of Cen A analogs are extremely useful for studying the impact of AGN on their host galaxy and the surrounding gas, relatively little work has been done on identifying their effects on satellite galaxies. \cite{Das&Pandey2025} explored star forming galaxies within $2\, \mathrm{Mpc}$ of AGN hosts in the EAGLE simulations. They found that 61\% of AGN adjacent star-forming galaxies are quenched and 39\% show enhancement in star formation, with quenching preferentially occurring in higher mass halos and enhancement in lower mass halos. A recent study by \cite{Visser-Zadvornyi+2025} argues that the interplay of AGN heating and the absence of gas accretion can exhaust the interstellar medium (ISM) and thereby quench satellites. The work of \cite{Dashyan+2019} explores the effects of AGN feedback from central galaxies on their satellites by comparing two sets of simulations (with and without AGN) of halos in the range $M_\mathrm{halo}=10^{12}$--$10^{13.4}\,\mathrm{M}_{\odot}$ at $z=0$. They find that AGN feedback begins to affect gas content as early as $z=2$ and that AGN feedback could increase the temperature and relative velocity of intergalactic gas, contributing to the quenching of satellites.
 
In \cite{Weerasooriya+2024}, we applied the star formation physics that reproduced the dwarf satellite population of the Milky Way and their star formation histories at $z=0$ to a Cen A analog using the \textsc{Galacticus} SAM \citep{Benson2012}. SAMs utilize a network of linked differential equations to approximate baryonic processes, enabling the simulation of galaxies evolving over cosmic time. This approach is effective in rapidly simulating galaxies, as demonstrated by previous studies \citep[e.g.][]{Henriques2009,BensonBower2010}, and allows the modeling of dwarf galaxies \citep{Weerasooriya+2023,Weerasooriya+2024,Ahvazi+2024,sommerville2020}. SAMs can incorporate baryonic physics into pre-existing N-body merger trees extracted from dark matter-only N-body simulations and into merger trees constructed using extended Press Schechter techniques. This approach offers a streamlined and practical means to simulate dwarf galaxies within a broader environmental context, and is particularly well-suited for testing a variety of physics scenarios and/or parameter space in galaxy formation models. Our model reproduces the overall satellite population of Cen A, the luminosity function within 700~kpc, as well as its luminosity--metallicity relation. This finding suggests that the physics governing star formation is similar regardless of host mass. However, the modeled Cen A luminosity function within a projected distance of $150$~kpc does not match well with the observations \citep{Weerasooriya+2024}, with the model overpredicting the median number of observed satellites at $M_\mathrm{V}\sim-15.8$ by a factor of $4\pm1$ and at $M_\mathrm{V}\sim-14$ by a factor of $\sim1.8-3.3$ (see appendix \ref{sec:stats}).

In this follow-up study, we use our SAM to study dwarf galaxies closest to the host. We will compare our simulations against an observational sample taken from the literature. We use a compilation of observational data for Cen A satellite dwarf galaxies within 700~kpc. These dwarfs were compiled from observations available through the Panoramic Imaging Survey of Cen \& Sculptor  \citep[PISCeS;][]{crnojevic2010,crnojevic2014,crnojevic2016,crnojevic2019}, the Survey of Cen A's Baryonic Structures \citep[SCABS;][]{SCABS,Taylor+2018}, Leahy et al. (private communication) and other studies by \cite{karachentsev2013,muller2015,muller2017,Muller2019}. The reader is referred to Table 1 of \cite{Weerasooriya+2024} for the complete data sample.

We focus on those satellites within a projected distance of 150~kpc of Cen A, as observational studies in this area are complete down to $M_\mathrm{V} \approx -10$. We explore this subset of dwarfs to test whether we can reproduce the correct shape of the observed luminosity function within $150\,\mathrm{kpc}$ by varying the strength of different physical processes or by including new AGN feedback physics. In section \ref{sec:sim}, we describe the sample of observational data taken from the literature and the details of the simulation and galaxy models. We further describe our motivation to investigate Cen A in section \ref{sec:motivation}. We explore the effects of tidal disruption in section \ref{sec:tidal} and quenching due to reionization in section \ref{sec:reion}. We then explore quenching due to heating of the intergalactic medium (IGM) in section \ref{sec:hotIGM} and discuss why this may be caused by AGN in section \ref{sec:clues}. Finally, we discuss and summarize our results in section \ref{sec:summary}.

% \section{Observational Sample}\label{sec:obs}

% Here, we describe a sample of observational data of Cen A taken from the literature to which we compare our models. 

\section{Simulations \& modeling}\label{sec:sim}

\subsection{Models based on N-body merger trees}\label{sec:NbodyTrees}
We apply the \textsc{Galacticus} SAM to merger trees extracted from a high-resolution cosmological N-body simulation of an isolated Cen A halo from \cite{bovill2016,Weerasooriya+2024}. This N-body simulation of a Cen A analog ($M\sim 1.67\times10^{13}\;\mathrm{M}_{\odot}$) uses WMAP9 cosmology \citep[$\sigma_8=0.821,\,H_0=70.0\, $km$\,$s$^{-1}\,$Mpc$^{-1},\,\Omega_b=0.0463,\,\Omega_{\Lambda}=0.721$,][]{Hinshaw2013ApJS..208...19H} and is run from $z\,=\,150$ to $z\,=\,0$ using a zoom-in technique. We generate the initial conditions for the simulation with \texttt{MUSIC} \citep{Hahn2011} and run the simulation using \texttt{Gadget 2} \citep{gadget}. We analyze the halo properties and explore their evolutionary history through \texttt{AMIGA} \citep{AHF} and \texttt{CONSISTENT\_TREES} \citep{consistent_trees} respectively. We approximate the local environment of Cen A using the isolation criterion that there be no halos $M\geq 10^{12}\;\mathrm{M}_{\odot}$ within $3\;$Mpc $\;h^{-1}$ of the Cen A analog at $z\,=\,0$. Then a higher resolution re-simulation is run, centered on the selected halo with effective number of particles $N_\mathrm{eff}=8192^3$, particle mass $\mathrm{m_p=1.4\times10^5\,\mathrm{M}_{\odot}}$ using a softening length $\mathrm{\epsilon=200\,h^{-1}\,pc}$.

While a few studies have looked at star formation rates of Cen A dwarfs \citep{cote+2009,crnojevic+2011}, the details of star formation physics in Cen A dwarfs are still largely unexplored. To begin our study, we assume that our galaxy models have the same stellar physics as the Milky Way satellite dwarf galaxies. Therefore, we use astrophysical prescriptions and parameters that reproduced the MW satellites in \citet{Weerasooriya+2023} and the overall Cen A satellite population \cite{Weerasooriya+2024} using \textsc{Galacticus}. These models employ reionization suppression physics, tidal stripping, ram pressure stripping, and stellar feedback. We apply these models to N-body trees of a Cen A analog, as these provide detailed spatial information, allowing us to explore the properties of dwarfs within a projected 2D distance of the host galaxy (as is done in observational studies).

This study focuses on the cumulative luminosity function of Cen A satellites. As our Cen A analog has no preferred viewing angle, we rotate the model distribution of galaxies around the $z$-axis of the simulation in $5$ degree steps and project the galaxy positions into the $y$–$z$ plane, selecting on projected radius at each step. This approach provides an estimate of the uncertainty in the predicted luminosity function due to the system's anisotropies and random fluctuations arising from the finite number of satellites.
    % and several realizations of Extended Press-Schechter (EPS) merger trees with positional information. 

\subsection{EPS merger trees}\label{sec:EPS}

In addition to the N-body merger tree used for the main analysis of this paper, we generate a statistical sample of merger trees to explore the effects of variation in the mass of the Cen A halo using the Extended Press-Schechter algorithm \citep[][see Appendix \ref{sec:massDep}]{Press&Schechter1974,Bond+1991,Bower1991,Lacey&Cole1993}. Our merger trees are generated using the merger tree building algorithm of \cite{Cole2000} with the modifier function of \cite{Parkinson+2008} and a minimum resolved halo mass $1.41\times10^7\,\mathrm{M_{\odot}}$, equivalent to 100 particles in the N-body simulation described above.

\section{Motivation: The curious case of Centaurus A} \label{sec:motivation}
\begin{figure*}[h!]
        \centering
       
          \includegraphics[width=\linewidth]{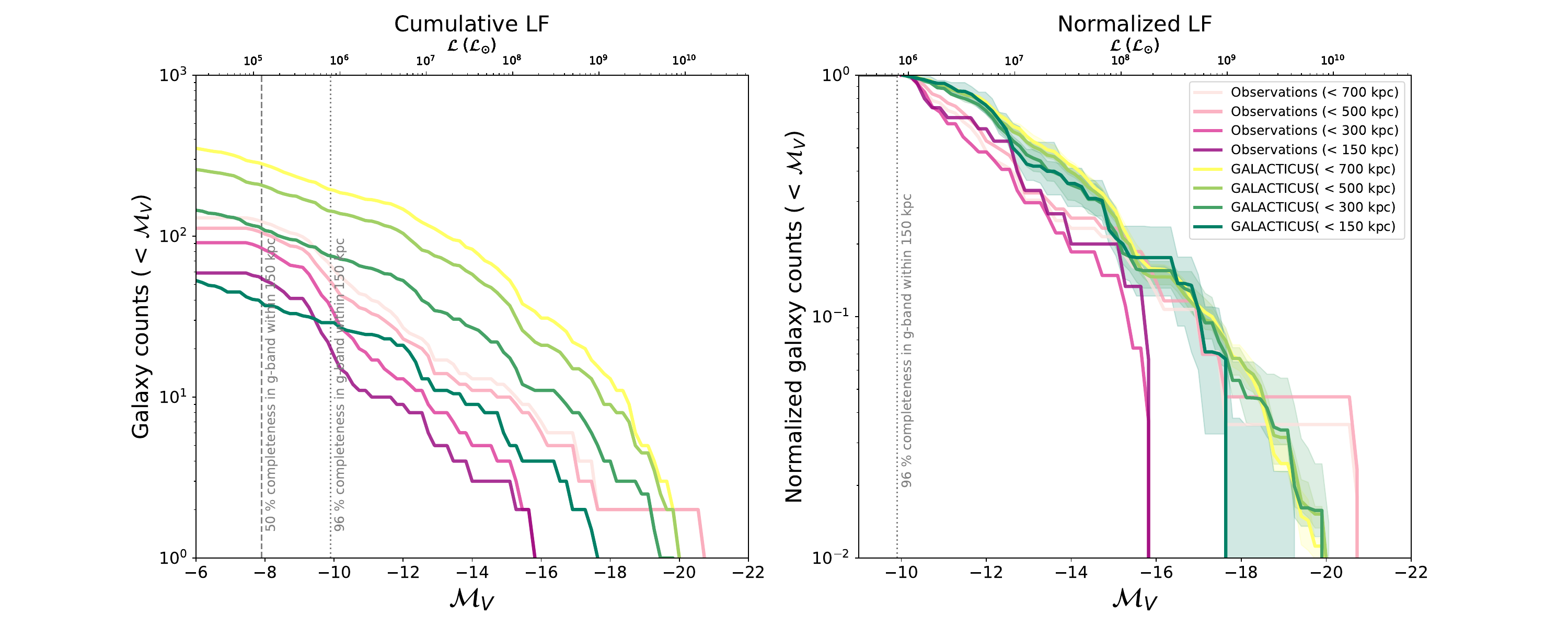}

        \caption{Left panel: Cumulative luminosity functions for Cen A satellites within 2D projected radii in the range $150$--$700\,\mathrm{kpc}$ as indicated by the legend. Green--yellow curves show the median number of satellites within each radius for our \textsc{Galacticus} models from \cite{Weerasooriya+2024}, while purple--pink lines show the same for the observations. The number of galaxies decreases as the projected radius decreases, due to the reduction in the sky area within that radius. In the observational data, there is a visible lack of luminous satellites with $M_\mathrm{V}<-15.8$ and an excess of fainter satellites $M_\mathrm{V}>-10$ (see the dark purple curve) compared to our model expectations (see the dark green curve). This lack of luminous satellites begins to emerge for luminosity functions within $300\,\mathrm{kpc}$. Right panel: Normalized cumulative luminosity functions for Cen A satellites. The shaded regions show the $1\sigma$ scatter in the number of galaxies in the predicted luminosity function due to anisotropies of the system (see section \ref{sec:sim}). The slope of the observed luminosity function begins to steepen as we move inward from $500$--$300\,\mathrm{kpc}$. The normalized cumulative luminosity functions show that models overpredict galaxy counts at all magnitudes, especially within 150 kpc, where observations brighter than $M_\mathrm{V}\leq-10$ are almost 100\% complete.}
        \label{fig:Mv_distance}
    \end{figure*}

\textsc{Galacticus} reproduces the luminosity function of the overall population of satellites within 700~kpc, but it overproduces the number of luminous satellites within 150~kpc \citep{Weerasooriya+2024} at all magnitudes. Despite the 96\% completeness in the observed luminosity function in this region (Leahy et al., private communication), there are very few bright satellites at all magnitudes and no satellites brighter than $-15.8\,\mathrm{mag}$. Therefore, in this work, we investigate the luminosity functions of Cen A satellites within $150$--$700\,\mathrm{kpc}$ projected distance for both observations and \textsc{Galacticus} models. In Figure~\ref{fig:Mv_distance}, we show luminosity functions of Cen A dwarfs within different projected distances from Cen A, ranging from $150$--$700\,\mathrm{kpc}$, for observations (in shades of pink/purple) and models (shades of yellow/green curves). 

As seen in \cite{Weerasooriya+2024}, the shapes of the modeled and observed luminosity functions agree quite well as we step inward from $700$ kpc to $500$ kpc, with the overall number of satellite galaxies decreasing with the area on the sky (see the left panel of Figure \ref{fig:Mv_distance}). Note that we have performed a 2 sample K-S test on normalized luminosity functions for samples $<500$ kpc and $<700$ kpc. Our null hypothesis is that two samples come from the same distribution. Comparison of normalized observed and modeled luminosity functions for $<500$ kpc reports a p value of 0.27 and $<700$ kpc reports a p value of 0.2. Therefore, we cannot reject null hypothesis for both cases. As we step into the inner regions, corresponding to smaller projected distances from Cen A, we see a similar trend of decreasing numbers of satellite galaxies for galaxies fainter than $M_\mathrm{V}\geq-15.8$.

However, we see an absence of observed luminous satellites ($M_\mathrm{V}\leq-15.8$) within projected radii $150$--$400\,\mathrm{kpc}$ and an excess of observed fainter satellites ($M_\mathrm{V}\geq -10$) relative to our model expectations. This lack of luminous satellites results in an increase in the slope of the luminosity function as we look within $500\,\mathrm{kpc}$ to $400\,\mathrm{kpc}$. Interestingly, within $150$--$300$~kpc, the observed luminosity function returns to its expected decline in normalization, due to the lower number of sub-halos found in these smaller projected areas, without any corresponding change in shape (see left panel of Figure \ref{fig:Mv_distance}). For a more careful analysis that accounts for incompleteness, we focus on an area within $150$~kpc of Cen A, as observations are 96\% complete to $M_\mathrm{V}\sim-10$ in this region (Leahy et al, private communication). This projected radius, $150$~kpc, is more appropriate for a nearly complete sample, as both the SCABS and PISCeS surveys covering Cen A detect satellites within this area.
In the right panel of Figure \ref{fig:Mv_distance}, we plot normalized cumulative luminosity functions (i.e., normalized to unity at the faintest magnitudes) to facilitate comparison of their shapes. Our model of Cen A from \cite{Weerasooriya+2024} overpredicts normalized galaxy counts at all magnitudes within 150~kpc, and produces satellites brighter than $M_\mathrm{V}\leq-15.8$, whereas observations lack these brighter satellites. The lack of luminous satellites within 150~kpc cannot be attributed to observational incompleteness, as these galaxies would be far brighter than even the 96\% observational completeness limit. In our \textsc{Galacticus} models, we do not see a similar trend of increasing slope as we look at luminosity functions within the inner regions of \textsc{Galacticus} (see right-panel curves shown in shades of green/yellow). Statistical analysis using the resampled distribution of galaxies from our fiducial model shows that a lower number of observed bright satellites is unlikely (see Appendix \ref{sec:stats} for details).    

This lack of luminous galaxies brings our attention to the curious case of Cen A satellites. Why are there so few bright satellites in the inner region of Cen A? One possibility is that this is the result of small--number statistics; perhaps Cen A is simply an outlier that randomly happens to have very few of these bright galaxies in its inner regions. The shaded regions in the right panel of Figure~\ref{fig:Mv_distance}, which show the $1\sigma$ uncertainties arising from anisotropies in the distribution of satellites, suggest that this is unlikely. In Appendix~\ref{sec:stats} we explore this question in more detail and demonstrate that statistical fluctuations are unlikely to explain this difference in the number of bright satellites that are predicted and observed.

We must therefore seek a physical reason for the dearth of bright satellites in the inner regions of Cen A. Were they significantly quenched making them much fainter, and if so, how did they quench? In the remainder of this paper, we test a variety of potential astrophysical mechanisms that may be responsible for the peculiarity of the cumulative luminosity function within $d~<~150$~kpc of Cen A at $z=0$. We explore several scenarios, including the effects of quenching mechanisms, different masses for Cen A analogs, and different merger histories (see Appendix \ref{sec:massDep}). The specific quenching pathways we explore are tidal stripping, tidal destruction, and reionization quenching. With the computational efficiency of \textsc{Galacticus}, we are able to determine which, if any, of these mechanisms can reproduce the observed quenching of dwarfs brighter than $M_\mathrm{V}~<~-16$ at $d~<~150$~kpc.

\section{Tidal Disruption}\label{sec:tidal}

The Cen A group of galaxies is richer and more massive than the Local Group, and so likely has a dynamically active environment in which tidal forces play a dominant role in the evolution of the group members. Here, we explore two consequences of these tidal forces on galaxies: (1) stripping of stars and gas in the ISM and (2) tidal destruction of halos.  

\subsection{Tidal Stripping of Interstellar Medium (ISM) and Stars}\label{subsec:tidal}

First, we look more closely at tidal stripping of satellite galaxies, implementing a simple model of ISM gas and stellar stripping. Stellar mass loss due to tidal stripping is given by (with an equivalent expression for ISM mass), 

\begin{equation}
    {\dot M_*}=\beta_\mathrm{tidal}\frac{F_\mathrm{tidal}}{F_\mathrm{res}}\frac{1}{T_\mathrm{dyn}}M_*,
\end{equation}
where $\beta_\mathrm{tidal}$ is a parameter controlling the efficiency of tidal stripping, $F_\mathrm{tidal}$ is the tidal force arising from the gravitational potential of the host halo (evaluated at the current orbital position of the satellite), $F_\mathrm{res}$ is the gravitational restoring force internal to the satellite (evaluated at its half-mass radius), $T_\mathrm{dyn}$ is the dynamical time at the half-mass radius of the galaxy, and $M_*$ is the stellar mass. Note that this model captures the effects of tidal stripping on the total mass and ignores any effect on the shape of the galaxy's density profile \citep{galacticus_}.

\begin{figure}[ht!]
    \centering
         \includegraphics[width=\linewidth]{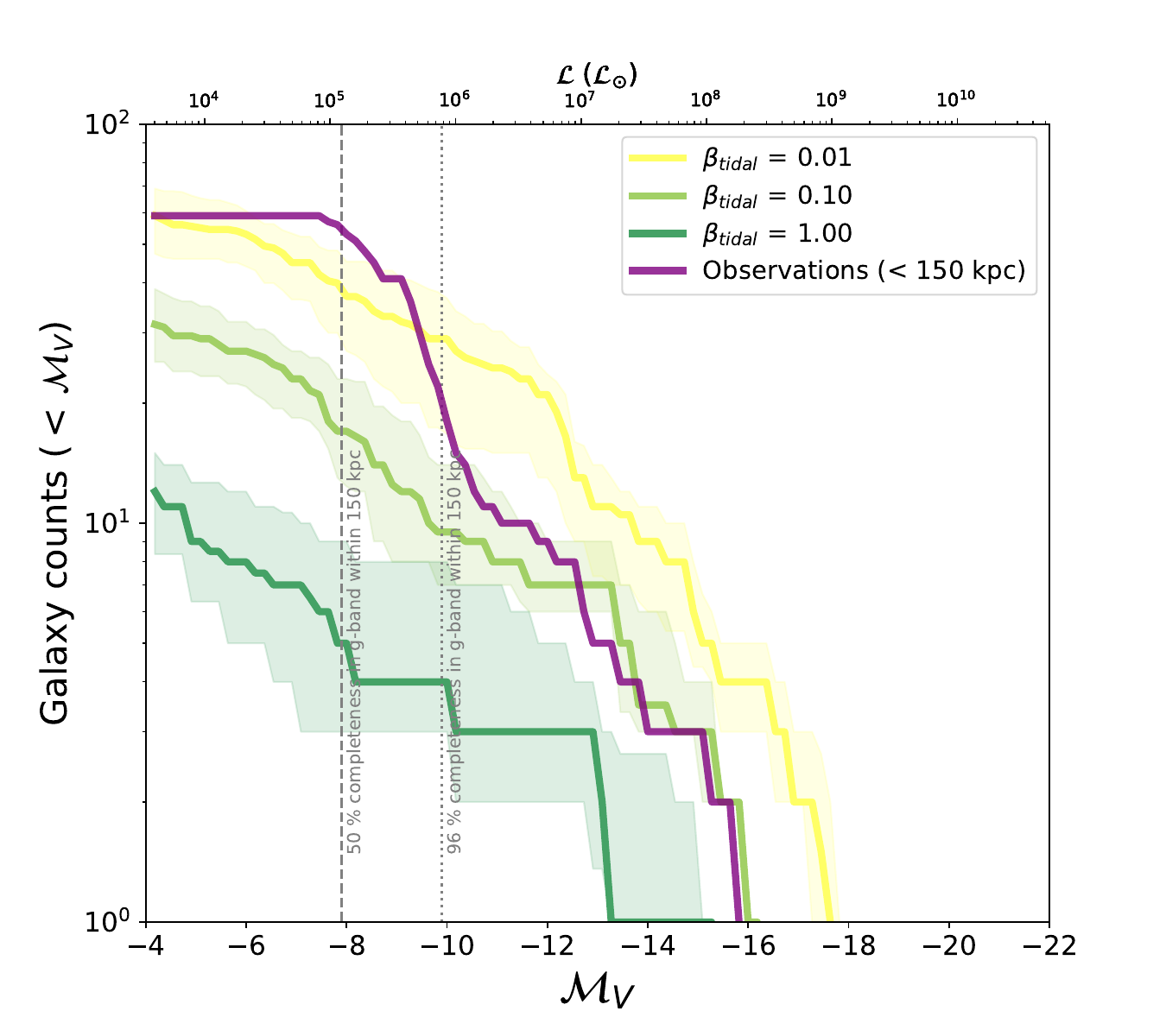} 
   \caption{Cumulative luminosity functions of Cen A satellites within a projected radius of 150~kpc. Observations are shown in purple, while green--yellow lines show results from our models with varying tidal stripping efficiencies from $\beta_\mathrm{tidal}=0.01$ (our default value) through to $1.0$, as indicated in the legend. }
   \label{fig:tidal}
\end{figure}

In Figure \ref{fig:tidal}, we explore the luminosity function of Cen A satellites within $150\,\mathrm{kpc}$ for various tidal stripping efficiencies compared to observations in the same region (shown by the purple line). Results are shown for $\beta_{tidal}=0.01$ (yellow, which we previously constrained to match the properties of MW satellites), 0.1 (light green), and 1.0 (dark green). As expected, the number of dwarfs at all luminosities decreases as the tidal stripping efficiency is increased; stronger tidal stripping removes existing stars and also ISM gas, leaving less available for star formation, resulting in fainter galaxies. The effect is similar to that found for the Milky Way satellites by \cite{Weerasooriya+2023}. Importantly, tidal stripping is not able to steepen the luminosity function. While it \emph{can} reduce the number of bright galaxies to match that observed (see the $\beta_\mathrm{tidal}$ line in Figure~\ref{fig:tidal}), this then results in too few faint galaxies. This is reasonable given that the tidal force depends only upon the host halo properties, and the internal gravitational forces in satellites do not vary strongly with galaxy mass.  Therefore, we can rule out tidal stripping as a possible cause of the steepening effect of the luminosity function within $150\,\mathrm{kpc}$.

\subsection{Tidal Destruction of Halos}

\begin{figure}[htbp!]
    \centering
    \includegraphics[width=\linewidth]{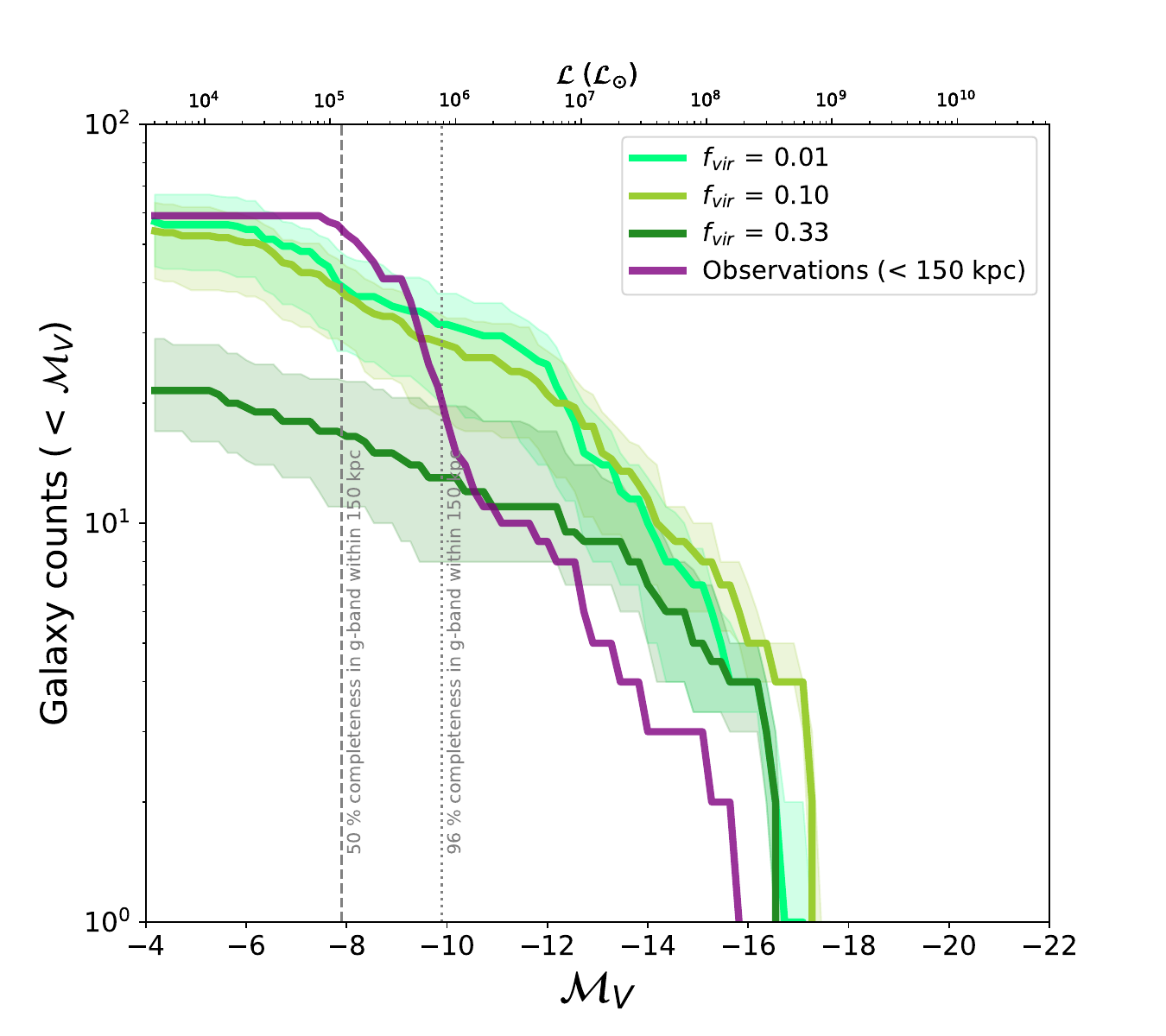}
    \caption{The cumulative luminosity function of Cen A satellites within a projected radius of $150\,\mathrm{kpc}$. Observations are shown by the purple line, with model predictions shown by the green--yellow lines. We present results for several models that include tidal destruction of subhalos across different virial fractions ($f_\mathrm{vir}$). Subhalos that pass within a 3D radius of $f_\mathrm{vir} r_\mathrm{vir}$ of Cen A are assumed to be destroyed due to the tidal field of Cen A.}
    \label{fig:tidal_des}
\end{figure}

In our standard model, we assume that tidal stripping and/or destruction of dark matter halos is accurately captured in the $N$-body simulation. However, our N-body simulation is of dark matter only. As such, it does not contain the central galaxy, Cen A itself. In reality, this central galaxy should strengthen the tidal field experienced by orbiting dark matter subhalos, potentially leading to an enhanced rate of mass loss and destruction.

To test if this enhanced tidal destruction could influence the number of bright satellite galaxies, we switched to using \textsc{Galacticus}'s analytic subhalo orbit model, as is typically employed when utilizing merger trees based on extended Press-Schechter theory \citep[see][for an application of this model to the Milky Way dwarf satellites, and for a description of the model itself]{Ahvazi+2024}. This model has the advantage that, unlike our dark-matter-only simulation, it includes the central galaxy's contribution to the tidal field. We apply this model unchanged from that used in \cite{Ahvazi+2024}, \emph{except} that subhalo orbits at infall are no longer drawn from a cosmological distribution, but are taken directly from the N-body merger tree. In this model, subhalos are considered destroyed if their bound mass falls below the minimum halo mass of the original N-body simulation ($1.4 \times 10^7\mathrm{M}_\odot$), or if their orbit passes within a 3D radius $f_\mathrm{vir} r_\mathrm{vir}$ of the central galaxy. Our default choice is $f_\mathrm{vir}=1\%$ \citep{Ahvazi+2024}. Here, we also explore $f_\mathrm{vir}=10\%$ and 33\% to test if enhanced tidal destruction could explain the observed dearth of bright galaxies.

Figure \ref{fig:tidal_des} shows the effect of tidal destruction on the cumulative luminosity function within the inner projected region ($<150\,\mathrm{kpc}$). For our default choice of $f_\mathrm{vir}=1\%$ (yellow line in Figure~\ref{fig:tidal_des}), our model using the analytic subhalo orbital model produces results broadly consistent with those from the N-body simulation, with the exception that it produces fewer of the very brightest satellites ($M_\mathrm{V} \lesssim -15.8$). This change may arise from differences in the effects of dynamical friction between our analytic orbit model and the N-body simulation, or could arise from small shifts in the phases of orbits (resulting in the subhalos hosting these most luminous satellites happening to lie outside of the 150~kpc region when evolved using our analytic orbit model). Nevertheless, the number of predicted galaxies brighter than $M_\mathrm{V}=-15.8$ remains substantially higher than that which is observed. As $f_\mathrm{vir}$ increases, the median number of fainter dwarfs decreases systematically, while the number of the brightest galaxies is affected much more weakly. For $f_\mathrm{vir}=10\%$, the number of the brightest galaxies is actually slightly increased (due to minor shifts in the phase of each subhalo along its orbit). For $f_\mathrm{vir}=33\%$, there is some decrease in the number of bright satellites,\footnote{Note that $f_\mathrm{vir} r_\mathrm{vir}$ is greater than 150~kpc in this case. As such, the only galaxies present in this luminosity function are those at 3D radii greater than $f_\mathrm{vir} r_\mathrm{vir}$ but which happen to lie at a projected radius less than 150~kpc.} but this is still not enough to bring the predictions into line with observations and comes at the expense of a greatly reduced number of faint galaxies, which is then in conflict with observations.

We conclude that enhanced tidal destruction from the baryonic components of the Cen A system is unable to reconcile our model with the observed number of bright satellites, even for very extreme choices for destruction (i.e., forcing all subhalos within 33\% of the virial radius to be destroyed).

\begin{figure}[ht]
    \centering
    \includegraphics[width=\linewidth]{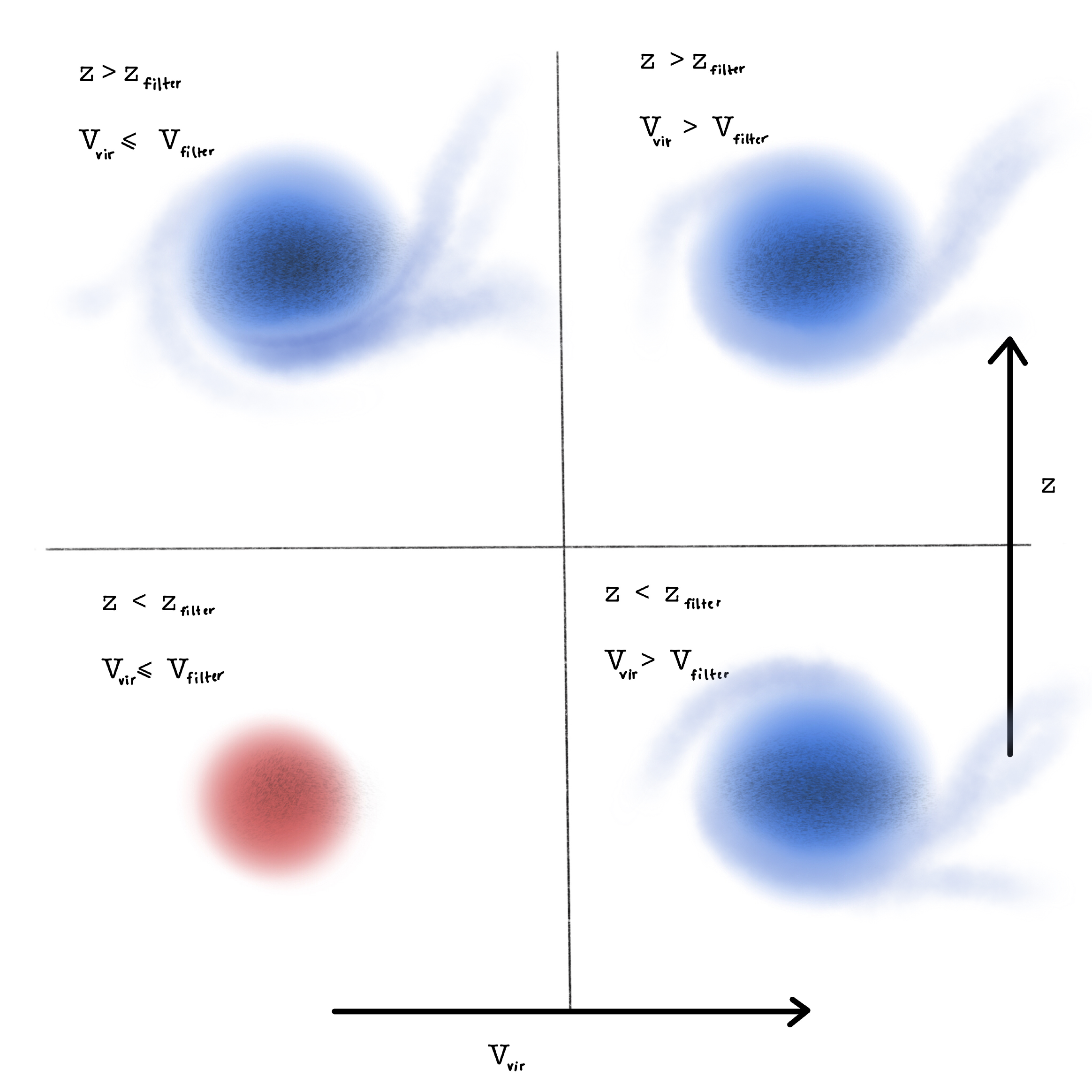}
    \caption{Reionization physics: This figure illustrates how quenching of star formation in dwarf galaxies due to reionization is approximated in \textsc{Galacticus}. Halos that accrete gas are shown in blue, while those that do not are shown in red. Gas accretion onto halos is suppressed when $z<z_\mathrm{reion}$ and $V_\mathrm{vir}<V_\mathrm{filter}$ (bottom left panel). We use $V_\mathrm{filter}=30\,\mathrm{km/s}$ \citep{RicottiGnedin:05} as gas accretion in galaxies with circular velocities $<20$--$30\,\mathrm{km/s}$ \citep{Bovill&ricotti2011} is suppressed by reionization.}
    \label{fig:reionization_vfilter}
\end{figure}

\section{Quenching due to Reionization}\label{sec:reion}

Reionization is the process by which the first stars reionized hydrogen throughout the Universe, resulting in the IGM being photoheated to temperatures of $\sim 10^4\,$K, which results in suppression of gas accretion into halos with shallow potential wells at later times. This process occurred over a period of around one billion years, from $z\,=\,14$--$6$  \citep{Fan+2006,Furlanetto+2006,Morales+2010}. The suppression of gas accretion onto halos depends both on the local redshift at which reionization occurs, $z_\mathrm{reion}$, and the local temperature of the IGM after reheating, $T_\mathrm{IGM}$. The relevant scale for this suppression is the so-called ``filtering mass'' \citep{RicottiGnedin:05} or, equivalently, filtering velocity, $V_\mathrm{filter}$. We use the reionization implementation used in \cite{Weerasooriya+2023} to model this process. In this implementation, accretion onto halos is suppressed when $V_\mathrm{vir}<V_\mathrm{filter}$, with $V_\mathrm{vir}$ being the virial velocity of the halo, which serves as a measure of the depth of the halo's gravitational potential well. Therefore, given a reionization redshift $z_\mathrm{reion}$, we suppress gas accretion to halos when $z<z_\mathrm{reion}$ and $V_\mathrm{vir}<V_\mathrm{filter}$.

Milky Way dwarfs are best fit by $z_\mathrm{reion}=9$ \citep{Weerasooriya+2023} and  $V_\mathrm{filter}=30\,\mathrm{km/s}$ \citep{RicottiGnedin:05}. \cite{Weerasooriya+2023} show that only dwarfs with $M_\mathrm{V}>-10$ are significantly quenched by reionization. However, Cen A may have reionized earlier and/or to a higher temperature than the average IGM due to its high virial mass, and potentially higher than average star formation and AGN activity.

\subsection{Impact of reionization redshift}

\begin{figure}[htbp]
    \centering
    \includegraphics[width=\linewidth]{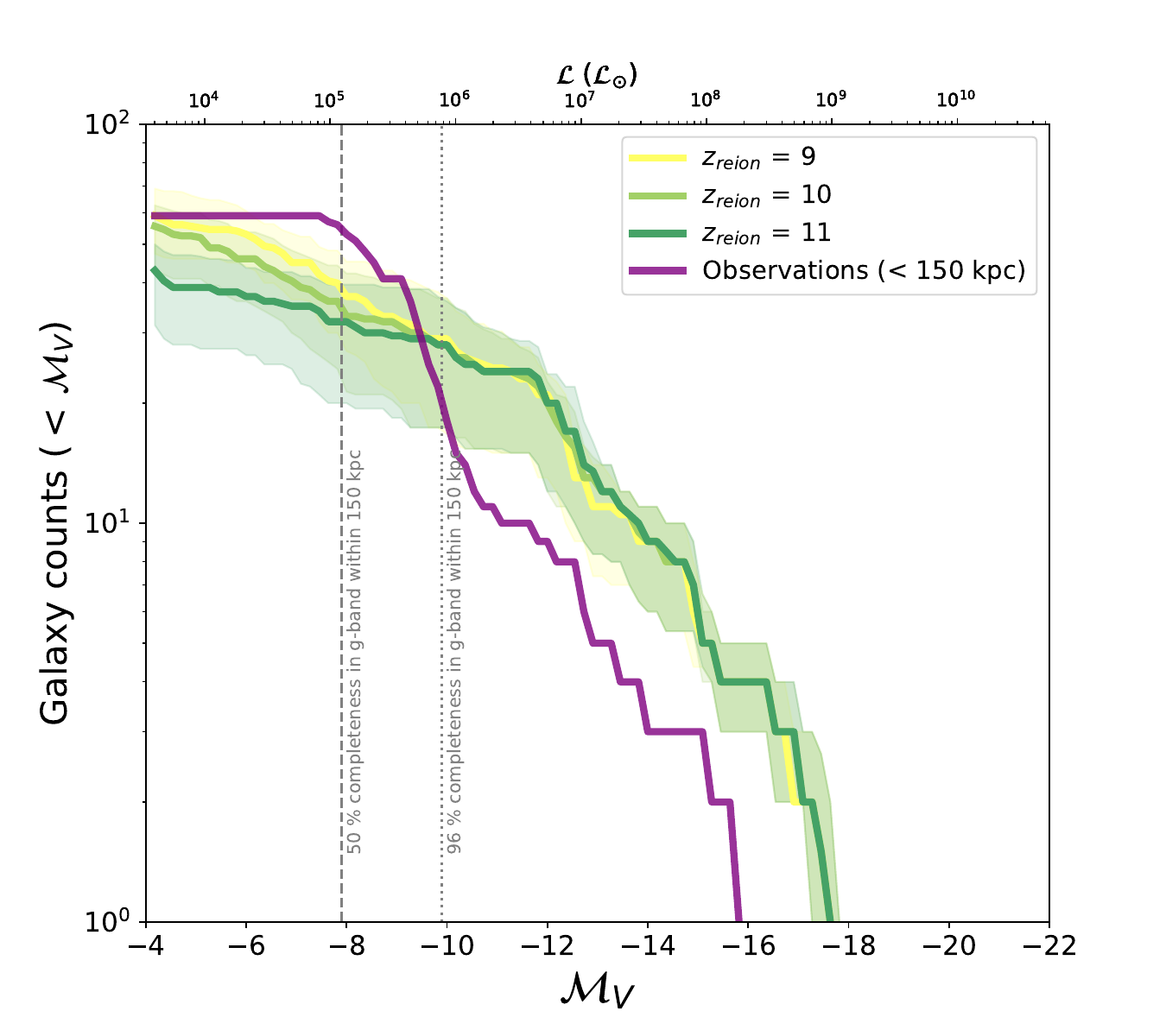}
    \caption{Cumulative luminosity functions for Cen A satellites within a projected radius of $150\,\mathrm{kpc}$. Observational data are shown by the purple line, with green--yellow lines showing model results. We show model predictions for three different reionization redshifts as indicated in the legend.}
    \label{fig:lum_Mv_z}
\end{figure}

Although our model reproduces the general population of Cen A satellites (within 700~kpc) at $z=0$ \citep{Weerasooriya+2024} assuming $z_\mathrm{reion}=9$, here we explore shifting reionization to higher redshifts (10, 11) to examine how Cen A satellites within $150\,\mathrm{kpc}$ are affected by earlier reionization.
Figure \ref{fig:lum_Mv_z} shows the cumulative luminosity functions for Cen A satellites within $150\,\mathrm{kpc}$ for both observations (purple) and \textsc{Galacticus} models (green--yellow). Increasing the redshift of reionization reduces the number of satellites with $M_\mathrm{V} > -10$. Note that the shaded envelopes in this figure represent the $1\sigma$ variation arising from the rotation of the modeled distribution of galaxies (as described in \S\ref{sec:NbodyTrees}). Even allowing for this range of variation, it is clear that changing the epoch of reionization can not reduce the number of bright satellites predicted by \textsc{Galacticus}. Therefore, we eliminate an earlier, localized redshift of reionization as a possible reason for the lack of bright Cen A satellites.

\subsection{Impact of higher temperature of IGM after reionization}

\begin{figure}[!ht]
    \centering
    \includegraphics[width=\linewidth]{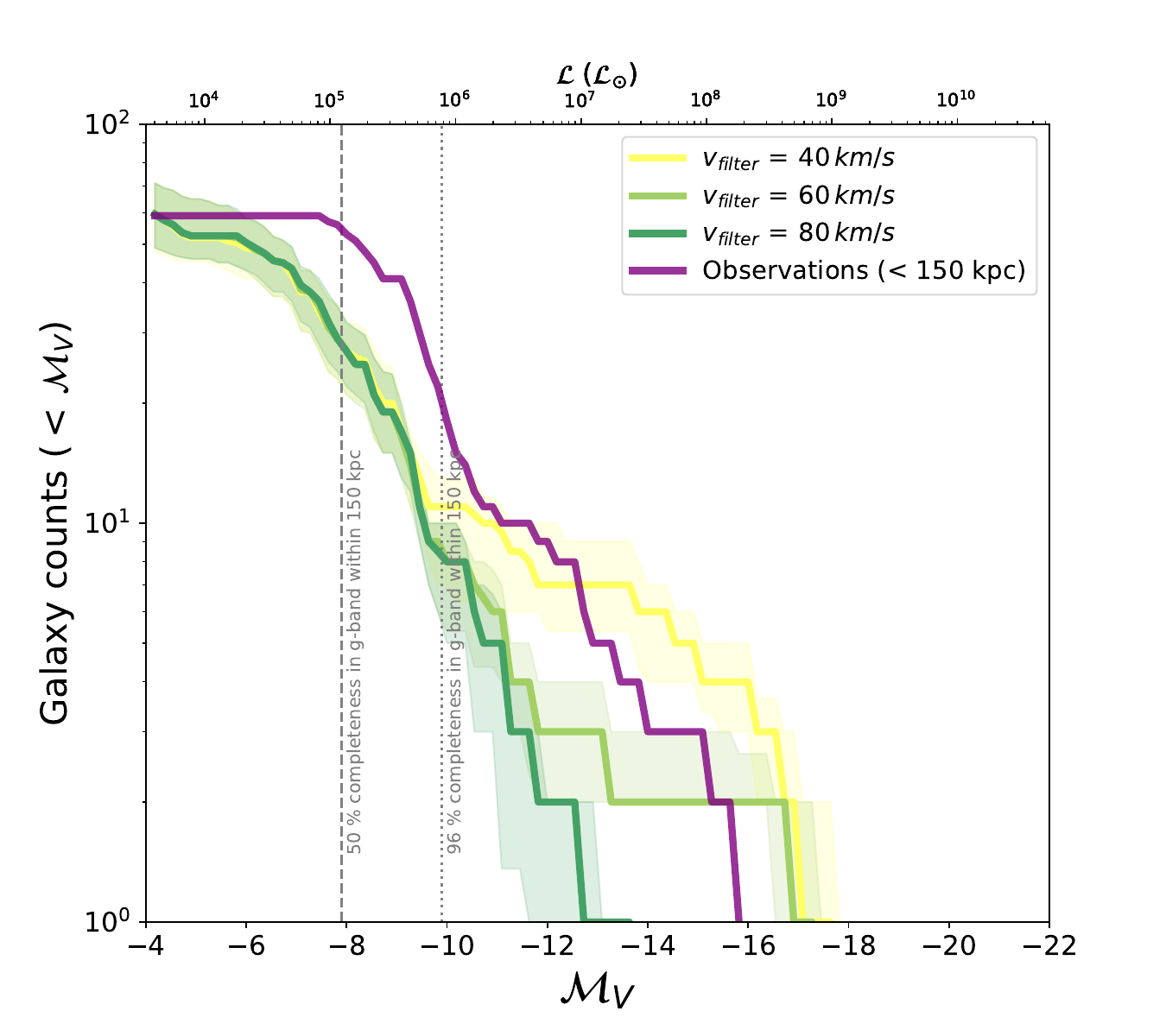}
    \caption{Cumulative luminosity functions of the Cen A satellites within $150\,\mathrm{kpc}$ in projection. Observations are shown by the purple line, with green--yellow lines showing model results. We model the effects of different post-reionization IGM temperatures in the vicinity of Cen A by varying the filtering velocity, $V_\mathrm{filter}$, to (40,60,80 km/s) as shown in the legend.}
    \label{fig:lum_Mv_vF}
\end{figure}

We now consider the possibility that Cen A was able to reheat its local IGM to a temperature significantly higher than that of the mean post-reionization IGM, and explore whether such a higher temperature could impact brighter satellites within the projected radius of Cen A $150\,\mathrm{kpc}$. To do so, we examine how the luminosity function changes as a function of the filtering velocity, $V_\mathrm{filter}$. In Figure \ref{fig:lum_Mv_vF}, we show the luminosity function within $150\,\mathrm{kpc}$ projected radius for $V_\mathrm{filter}=40, 60$, and $80\, \mathrm{km\,s^{-1}}$. As the filtering velocity increases, the slope of the luminosity function steepens due to a substantial decrease in the number of luminous satellites. 

To quantify the comparison between models and observations, we compare the luminosity functions within $150 \,\mathrm{kpc}$ using a 2-sample K-S test to determine if they come from the same distribution. This tests only the shape (slope) of the luminosity function, not the overall normalization. If the $p$-value is less than 0.05, we reject the null hypothesis that two samples come from the same distribution. We find the following $p$ values; 0.009 for $V_\mathrm{filter}=40\;\mathrm{km}\,\mathrm{s}^{-1}$, 0.02 for $V_\mathrm{filter}=60\;\mathrm{km}\,\mathrm{s}^{-1}$, and 0.57 for $V_\mathrm{filter}=80\;\mathrm{km}\,\mathrm{s}^{-1}$. Since the $p$-value of $V_\mathrm{filter}=80\;\mathrm{km}\,\mathrm{s}^{-1}$ is greater than 0.05, we cannot reject the null hypothesis in this case. While the slope of the luminosity function for $V_\mathrm{filter}\,=\,80\,\mathrm{km\,s^{-1}}$ matches that observed for the Cen A satellites with $d<150\,\mathrm{kpc}$ quite well, this strong suppression results in too few galaxies at all luminosities for which the observational sample is complete.
   
A filtering velocity of $80\,\mathrm{km\,s^{-1}}$ corresponds to an IGM temperature of $\sim 230,000\,K$, which cannot be attained by photoheating from star formation (typically limited to temperatures of order 20,000~K). However, such temperatures are reasonable in the warm-hot intergalactic medium (WHIM), and could potentially be reached if, for example, AGN activity were responsible for ionizing the region around Cen A.  Thus, only halos with masses $\geq 2\times 10^9 (z\sim9)-4.3\times10^8(z\sim5)\,\mathrm{M}{\odot}$ could accrete gas from the IGM, with strongly suppressed accretion onto lower mass halos. Higher IGM temperature leads to a higher filtering mass/filtering velocity and results in suppression of halos at higher luminosity. In the next sections, we investigate what happens if we reheat the IGM to a higher temperature at early redshifts.

\section{Role of combined effects}

In previous sections, we consider the role of individual quenching mechanisms such as tidal stripping, reionization, etc. on dwarfs of Cen A. While these did not reproduce the desired shape of the observed luminosity function, they may still play a role in combination. We test combined effects of tidal stripping + early reionization, and higher filtering velocity + early reionization to test additional scenarios.\\

\begin{figure}[!ht]
    \centering
    \includegraphics[width=\linewidth]{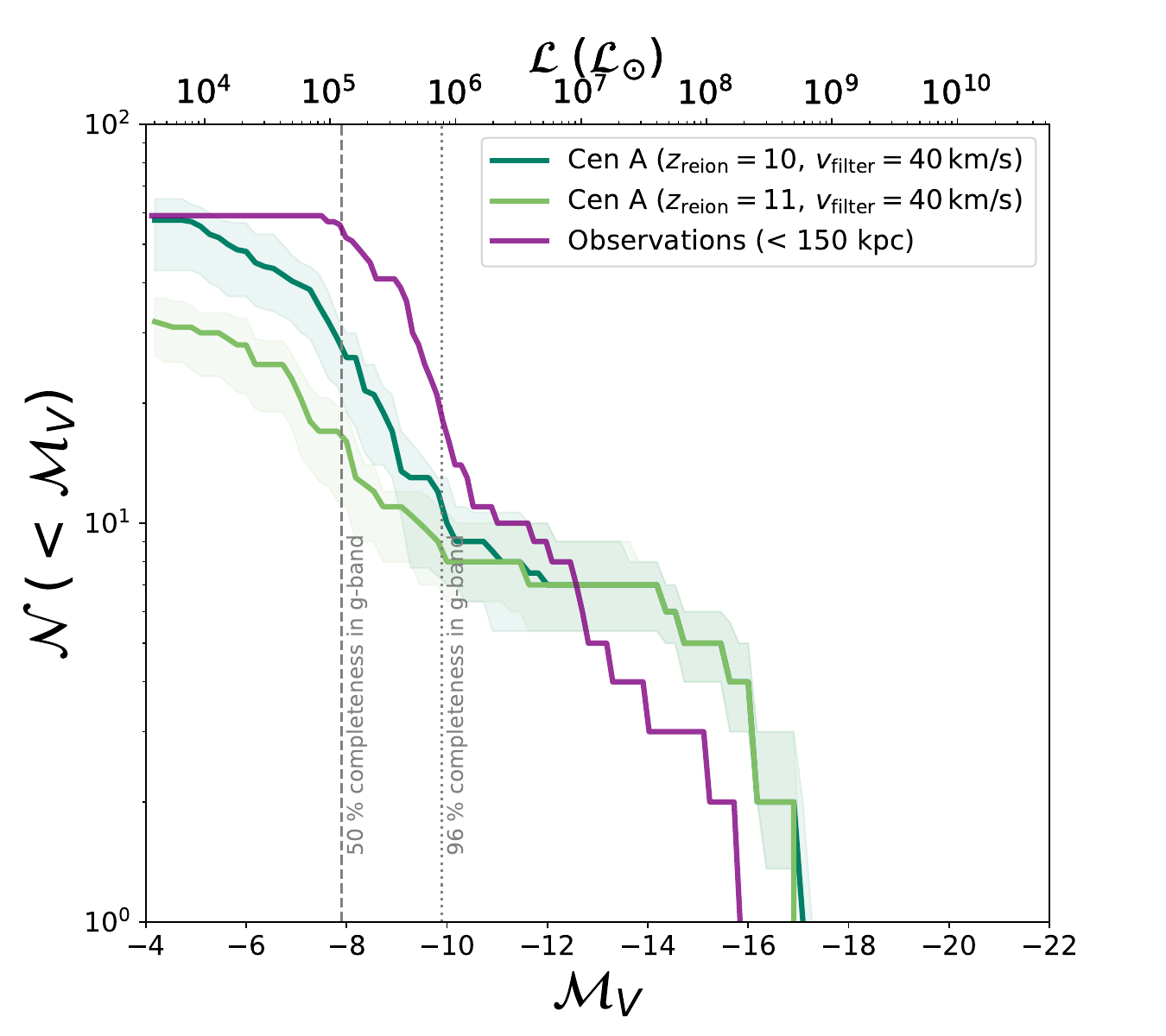}
    \caption{Cumulative luminosity functions of the Cen A satellites within $150\,\mathrm{kpc}$ in projection. Observations are shown by the purple line, with green--yellow lines showing model results. We model the combined effects of higher post-reionization IGM temperatures in the vicinity of Cen A by varying the filtering velocity, $V_\mathrm{filter}$, to 40 km/s from 25 km/s together with higher reionization redshifts as shown in the legend.}
    \label{fig:lum_Mv_reion_combined}
\end{figure}

Figure \ref{fig:lum_Mv_reion_combined} shows the combined effects of higher post reionization IGM temperature for 40 km/s along with reionization redshifts of 10 (green) and 11 (light green) respectively. However, earlier reionization suppress fainter satellites e.g. underpredict the number of satellites at $\mathrm{M_V}~-8$ (see light green curve). Thus, combined effects of reionization still do not reproduce the observed luminosity function within $150\,\mathrm{kpc}$ of Cen A.

\begin{figure}[!ht]
    \centering
    \includegraphics[width=\linewidth]{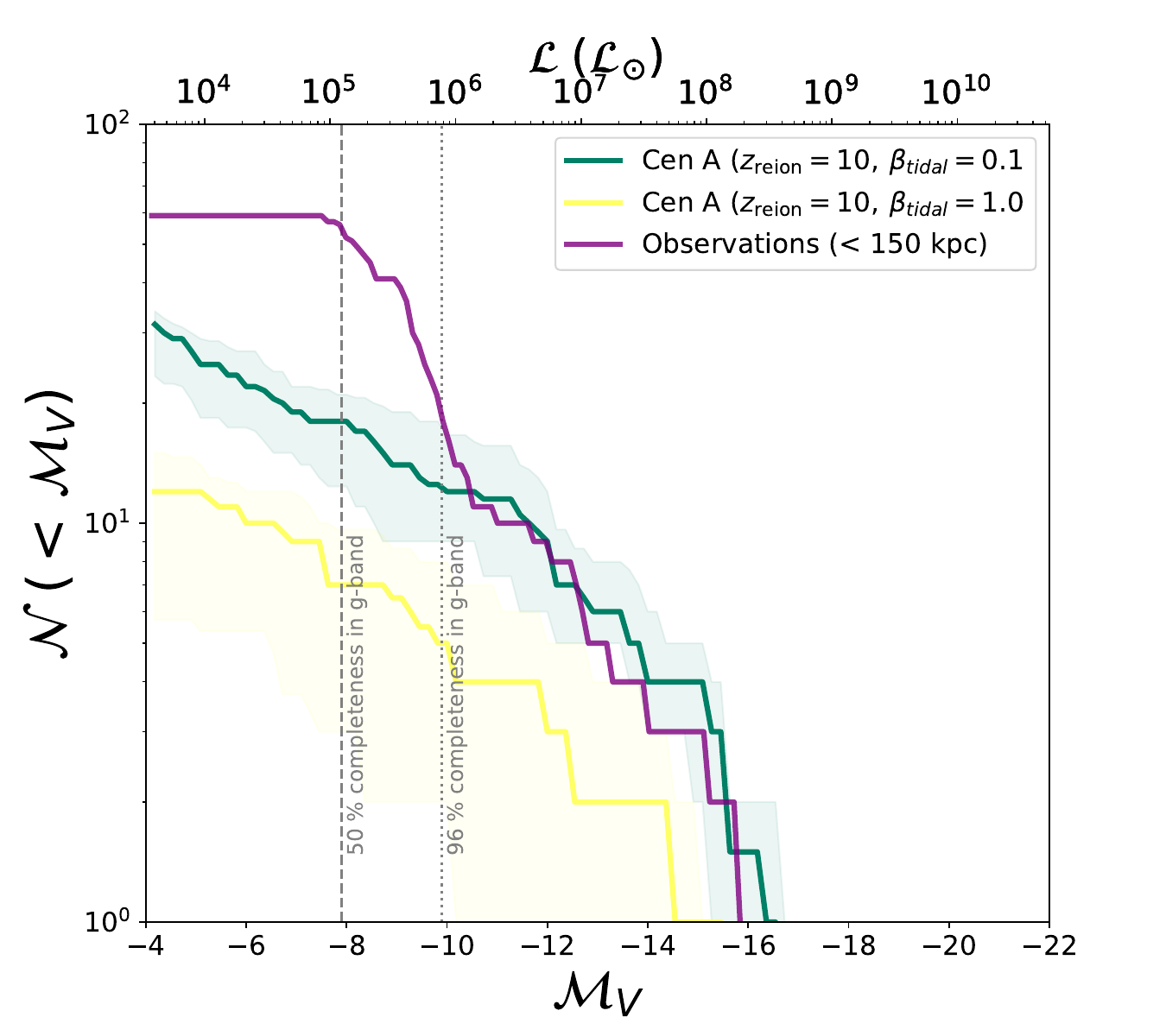}
    \caption{Cumulative luminosity functions of the Cen A satellites within $150\,\mathrm{kpc}$ in projection. Observations are shown by the purple line, with green--yellow lines showing model results. We model the combined effects of earlier reionization of Cen A and higher strength of tidal stripping.}
    \label{fig:lum_Mv_reion_tidal}
\end{figure}

Next, we test the combined effects of earlier reionization and tidal stripping (see Figure \ref{fig:lum_Mv_reion_tidal}). Increasing the strength of tidal stripping efficiency from 0.01 to 0.1 with earlier reionization reproduces the luminosity function within $-16\leq\,\mathrm{M_V}\leq-10$. However, fainter satellites at $\mathrm{M_V}\sim-8$ are under-predcited. Increasing the tidal stripping strength to 1 decreases the number of satellites at all luminosities compared to observations. These results further show that combined effects effects of reionization and tidal stripping cannot reproduce the observed luminosity function within 150 kpc. 

\section{Effects of hotter IGM}\label{sec:hotIGM}

% \section{Can higher IGM temperature quench star formation of Cen A satellites?}\label{sec:quenching}
We now investigate the possibility of a higher IGM temperature quenching the star formation of Cen A satellites. A substantially higher IGM temperature or, equivalently, a higher suppression velocity was not reasonable at the epoch of reionization. However, suppression of accretion into subhalos as a result of higher local IGM temperature at a lower redshift remains a possibility. The accumulation of gas in a halo is suppressed if the IGM temperature is greater than the virial temperature of the halo ($T_\mathrm{IGM}\geq T_\mathrm{vir}$), where the virial temperature is given by $T_\mathrm{vir}= \mu\,m_\mathrm{p}\,V_\mathrm{vir}^2 / 2\mathrm{k}_\mathrm{B} =35.9\left(V_\mathrm{vir}^2 / \hbox{km}\,\hbox{s}^{-1}\right)$. We can approximate this suppression of gas accretion by introducing a quenching velocity $V_\mathrm{quench}^2\propto\,T_\mathrm{IGM}$ into the model. Therefore, we extend the current gas accretion model in \textsc{Galacticus} to implement a quench velocity $V_\mathrm{quench}$, where the accretion of gas from the IGM is disabled for halos with circular velocities less than the quench velocity ($V_\mathrm{vir}<V_\mathrm{quench}$) below a given redshift ($z_\mathrm{quench}$), with accretion following the original model at earlier redshifts and for halos above the quenching velocity. Therefore, no gas is accreted into these halos after $z_\mathrm{quench}$. Any previously accreted gas in these halos can continue to fuel star formation until it is depleted.
    
% $$\mathrm{\dot{M}_{accretion}}=(\Omega_b/\Omega_m)\,\mathrm{\dot{M}_{halo}}$$
%\[
%\dot{M}_{\mathrm{accretion}} =
%\begin{cases}
%\dfrac{\Omega_b}{\Omega_m}\,\dot{M}_{\mathrm{halo}}, &
%V_{\mathrm{vir}} > V_{\mathrm{quench}}\
% \text{and}\
%z > z_{\mathrm{quench}} \\
%0, & \text{otherwise.}
%\end{cases}
%\]

% \begin{figure*}[!ht]
% \begin{center}
%     \begin{tabular}{cc}
% \includegraphics[width=.3\textwidth]{accretion_.pdf}
% \includegraphics[width=.3\textwidth]{accretion0.pdf}
%     \end{tabular}
%    \end{center}
%     \caption{Schematic illustration (for visualization purposes only; not direct simulation results) showing two modes of accretion impacted by IGM temperature or $V_\mathrm{quench}$. Gas accretes for halos with $V_\mathrm{vir}>V_\mathrm{quench}$ (left panel). No gas is accreted into these halos after $z_\mathrm{quench}$ (right panel). We set accretion to zero for halos with circular velocities less than the quench velocity ($V_\mathrm{vir}<V_\mathrm{quench}$, right panel) at a given $z\leq z_\mathrm{quench}$. Note that this is in addition to the accretion suppression due to reionization.}
%     \label{fig:TIGM}
% \end{figure*}

\subsection{When do luminous dwarf galaxies quench? }\label{sec:quench_redshift}

\begin{figure}[!ht]
    \centering
\includegraphics[width=\linewidth]{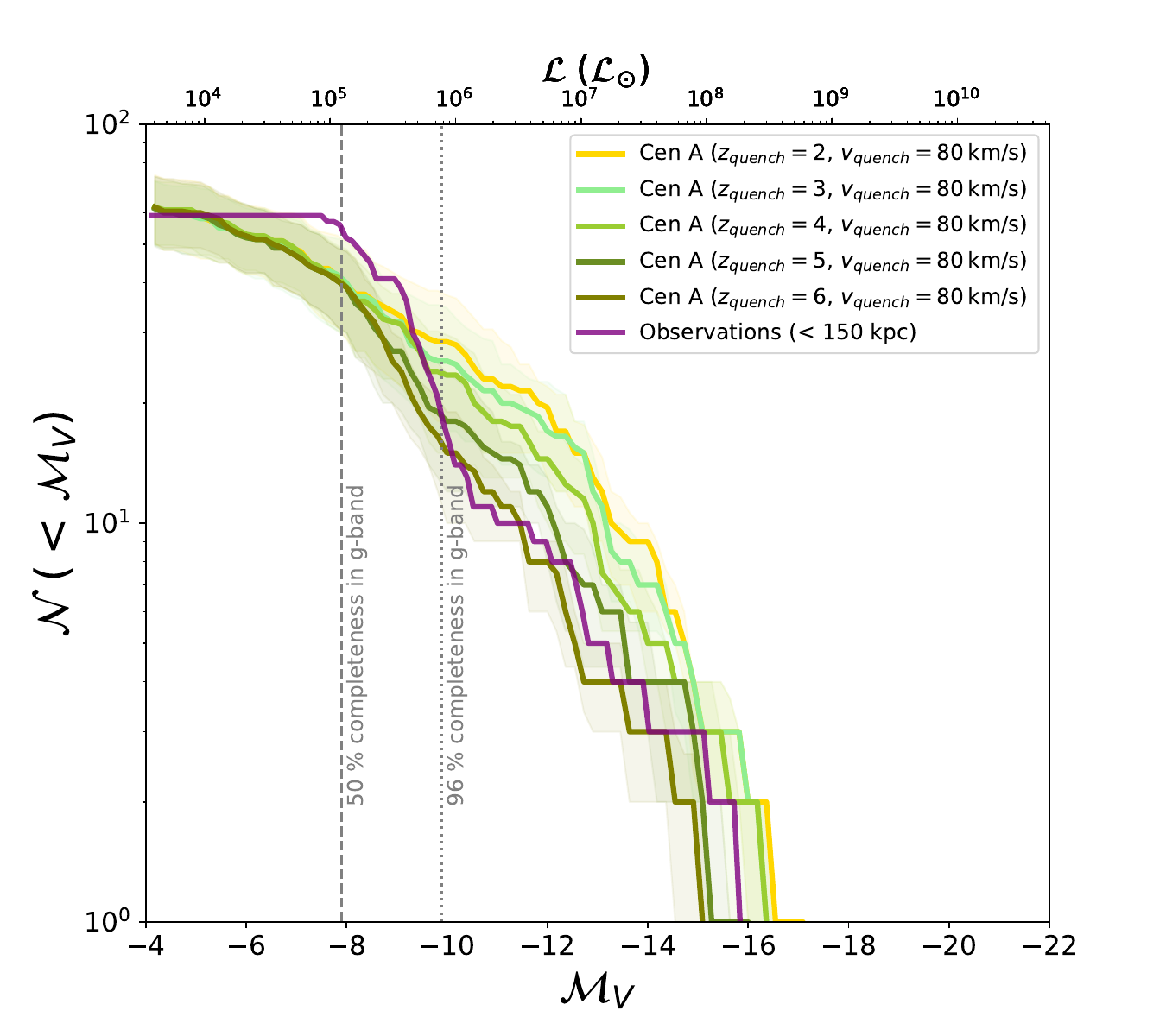} 
    \caption{Luminosity functions for the Cen A satellites within $150\,\mathrm{kpc}$ for galaxies quenched at $z_\mathrm{quench}=2-6$ with $V_\mathrm{quench}=80\,\mathrm{km/s}$. The purple line shows the observations within $150\,\mathrm{kpc}$ and \textsc{Galacticus} models are shown in shades of yellow/green.}
    \label{fig:lum_Mv_zquench80}
\end{figure}

Next, we heat the IGM at different redshifts to test how this affects the luminosity function. Using the filtering velocity found in Figure \ref{fig:lum_Mv_zquench80}, we run a small grid of models with an IGM temperature of $\sim 229,760\,\mathrm{K}$, equivalent to $V_\mathrm{quench}=80\,\mathrm{km/s}$ at redshifts 2, 3, 4, 5 and 6. Figure \ref{fig:lum_Mv_zquench80} shows the cumulative luminosity function for these models within a $150\,\mathrm{kpc}$ projected radius. We find that for $V_\mathrm{quench}=80\,\mathrm{km/s}$, the slope of the luminosity function steepens as the redshift of quenching increases from 2 to 6, with a quenching redshift of 5 giving the best match to the luminosity function. This suggests that Cen A may have prevented accretion into halos as early as $z=5$.

\begin{figure}
    \centering
    \includegraphics[width=\linewidth]{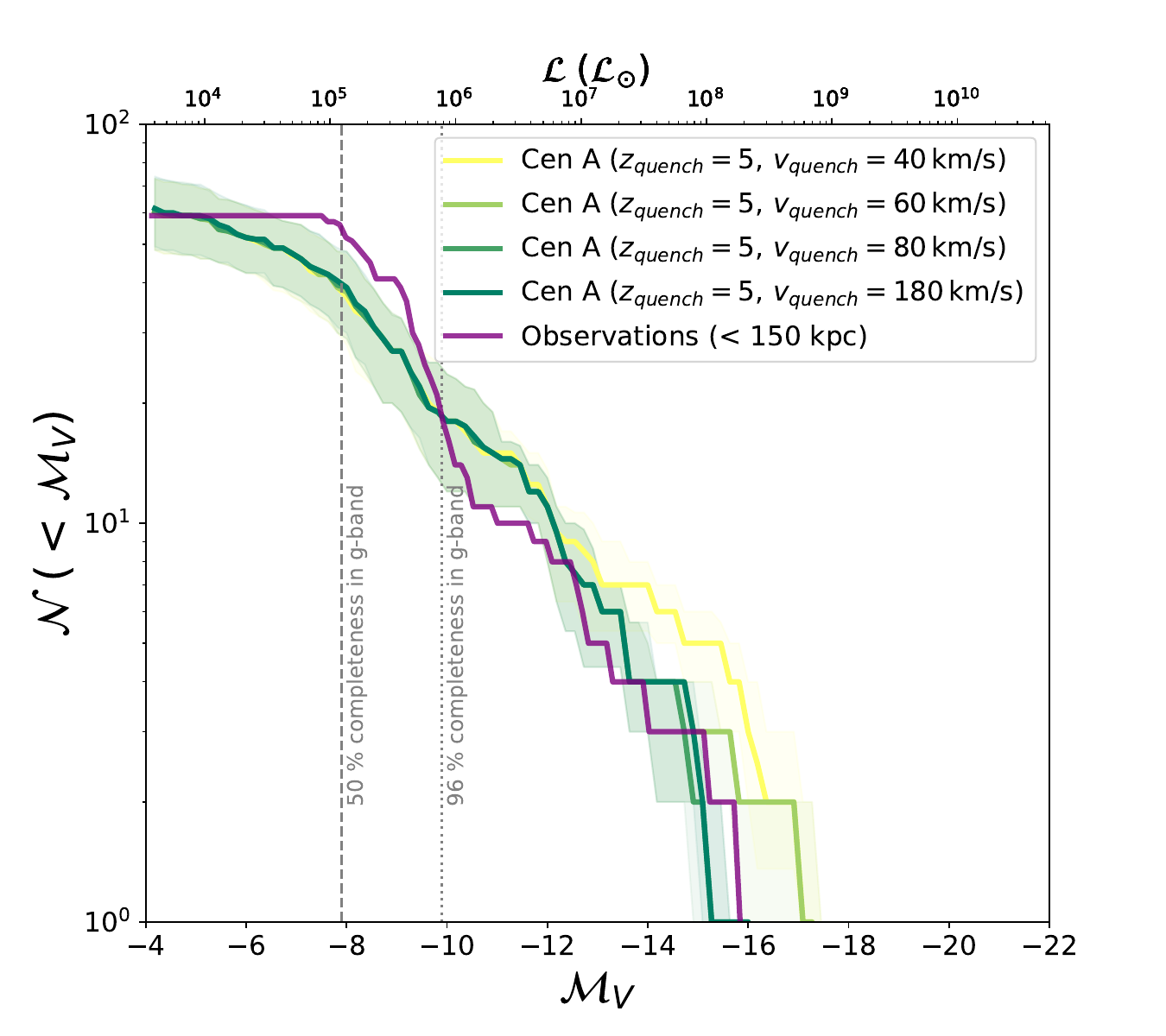}
    \caption{Cumulative luminosity functions for \textsc{Galacticus} models within $150\,\mathrm{kpc}$ for varying IGM temperatures (or $V_\mathrm{quench}=40,60,80,180\,\mathrm{km/s}$) at $z=5$ shown in light yellow, green, and blue respectively. The observed luminosity function is shown in purple. As IGM temperature or quench velocity increases, more luminous satellites ($M_\mathrm{V}\leq-15.8$) are quenched.}
    \label{fig:Mv_z5_vquench}
\end{figure}

\subsection{Quenching due to hot IGM at early redshifts}

 Next, we keep the quenching redshift at $z_\mathrm{quench}=5$ and vary the quenching velocity from $V_\mathrm{quench}=40,60,80,180\,\mathrm{km/s}$ (see Figure \ref{fig:Mv_z5_vquench}), corresponding to virial temperatures of $57,440$, $1.29\times10^3$, and $2.29\times10^5,1.16\times10^6$ K, respectively. Comparison of these models shows good agreement for $V_\mathrm{quench}=80\,\mathrm{km/s}$.
 
 What could have made the IGM around Cen A dwarf satellites so hot? Cen A is known to host an AGN with radio lobe temperature of $0.5\,\mathrm{eV}\approx5.80\times 10^6\,\mathrm{K}$ \citep{OSullivan+2013,Stawarz+2013}, comparable to the temperature necessary to facilitate suppression of gas accretion and quenching of satellite galaxies. Therefore, a possible explanation for this quenching effect is the heating of the IGM due to AGN activity.

\section{Following clues: AGN physics}\label{sec:clues}

 Our exploration of parameter space hints at quenching of dwarf galaxies due to higher IGM temperature.

\subsection{Area influenced by AGN}

\citet[][in prep]{Vazquez+2025} has shown that $\textsc{Galacticus}$ galaxies can reasonably reproduce distributions of black hole accretion rates (i.e. their Eddington ratios), and the $M$--$\sigma$ relation (see Figure \ref{fig:Msigma}) with minimal calibration. Therefore, we use \textsc{Galacticus} predictions to calculate BH properties at $z=5$, and to investigate the region influenced by AGN. In this section, we address the question, ``Are dwarf galaxies forming around Cen A's progenitor within the AGN's radius of influence at $z=5$?"

 To answer this, we first calculate the bolometric luminosity of Cen A's AGN in our model using the predicted accretion rate $\dot{M}_\bullet$ and radiative efficiency, $\epsilon$, of the black hole ($\epsilon\dot{M}_\bullet \mathrm{c}^2$, where $\mathrm{c}$ is the speed of light) at $z=5$. We find this value to be $6.02\times10^{42}\,\mathrm{erg/s}$. We approximate the ionizing luminosity using the X-ray bolometric correction factor of 12.83 \citep{Duras+2020}, which gives $L_\mathrm{ionizing}=L_\mathrm{bolometric}/12.83=4.69\times10^{41}\,\mathrm{erg/s}$. The radius of influence of the AGN is dependent on the bolometric luminosity of the AGN and the redshift. We derive the radius of influence by equating the total number of ionizing photons emitted by the AGN with the total recombination rate within the ionized region, assuming photoionization equilibrium (i.e. a Str\"{o}mgren sphere approximation. Note that in reality, energy injected from the AGN may not be isotropic and the equilibrium radius we compute is effectively an upper bound on radius to which the AGN may have ionized its surroundings - if the AGN has switched on only recently the actual ionized volume will still be growing, and therefore smaller in extent):

 \begin{equation}
 \frac{L_\mathrm{ionizing}}{E_\mathrm{H}}=\frac{\epsilon\dot{M}_\bullet \mathrm{c}^2}{E_\mathrm{H}}=\int_0^{R_{\mathrm{influence}}} 4\pi\,r^2\,n(r)^2\,\alpha_H\,\mathrm{d}r,
 \label{eq:Rinfluence}
% $$R_{\mathrm{influence}}={\bigg({\frac{3\epsilon\dot{M}c^2}{4\pi\,E_H\,\alpha_H\,n_H^2}}}\bigg)^{1/3}=\int_0^{R_{influence}}4\pi^r^2\,\rho(r)^2\,\alpha_H$$
\end{equation}

where $E_\mathrm{H}=13.6\,\mathrm{eV}=2.1\times10^{-11}\,\mathrm{ergs}$ is the ionization energy of hydrogen, $n(r)$ is the number density of hydrogen atoms as a function of radius around Cen A, and $\alpha_\mathrm{H}$ is the recombination rate coefficient for hydrogen. 

To model the distribution of gas around Cen A, we test different gas density profiles, including the gas density profile of \cite{Sorini+2024} and the gas density profile for poor group galaxies \citep{Mulchaey1998,Sun2012}. Furthermore, we add a background density equal to the mean density of gas in the IGM at large radii. The gas density profile for \cite{Sorini+2024} is given by:

\begin{equation}
 \rho(r)=\rho_{R}\,\bigg[\bigg(\frac{r}{R_{200}}\bigg)^2+\bigg(\frac{r_c}{R_{200}}\bigg)^2\bigg]^{\eta/2}+\rho_\mathrm{mean},
 \label{eq:sorini}
\end{equation}
where $r_\mathrm{c}=0.05\,R_{200}$, $\rho_R$ is a normalization factor and $\rho_\mathrm{mean}(z)=\Omega_\mathrm{b}\,\rho_\mathrm{c,0}\,(1+z)^3$ is the mean density of the universe, evaluated at $z=5$. 
Given that these simulations cannot resolve the density profile below the radii of $0.05R_{200}$ (where $R_{200}$ is the virial radius of Cen A defined as the radius enclosing a mean density equal to 200 times the critical density) we introduced a core into this profile.\\
For the Cen A progenitor at $z=5$ in our model applied to our N-body simulation we find that the main progenitor of Cen A has $R_{200}=41.88\,\mathrm{kpc}$ (physical) and $\log_{10}(M_\mathrm{halo}/\mathrm{M}_\odot)=11.6$ at $z=5$. We therefore adopt the following values from \cite{Sorini+2024}, $\eta=2.117$, $\log A_0=0.3$, $\alpha=0.07$, $\beta=-0.07$, $\eta_0=2.4$, from which we derive $\rho_R(z=5)=10^{\log A}\,\rho_\mathrm{c}(z=0)(1+z)^3$, where $\log A=\log A_0+\alpha \log(1+z)=0.354$. We assume a temperature of 229,760~K (corresponding to $V_\mathrm{quench}=80\,\mathrm{km/s}$), resulting in a recombination rate coefficient of $\alpha_H=1.356\times10^{-14}\,\mathrm{cm^3\,s^{-1}}$. 
  
 % Critical density at z=5 is $$\rho_c(z=5)=\rho_{c,0}(\Omega_m(1+z)^3+\Omega_{\Lambda})=9.2\times10^{-30}((0.28\times216)+0.721)\,\mathrm{g/cm^{-3}}=5.63\times10^{-28}\mathrm{g/cm^{-3}}$$
 
This density is then converted to the number of hydrogen atoms per unit volume: $n(r)=\rho(r) X_\mathrm{H}/m_\mathrm{H}$, where $m_\mathrm{H}$ is the mass of the hydrogen atom, and $X\approx0.75$ is the mass fraction of hydrogen.\\

We also test the gas density profile of poor group galaxies given by \citep{Mulchaey1998,Sun2012}:
\begin{equation}
    \rho(r)=\rho_0(r/r_0)^{-1.5}+\rho_\mathrm{mean},
\end{equation}
where $r_0=10\,\mathrm{kpc}$.
 \begin{figure}[h!]
     \centering
     \includegraphics[width=\linewidth]{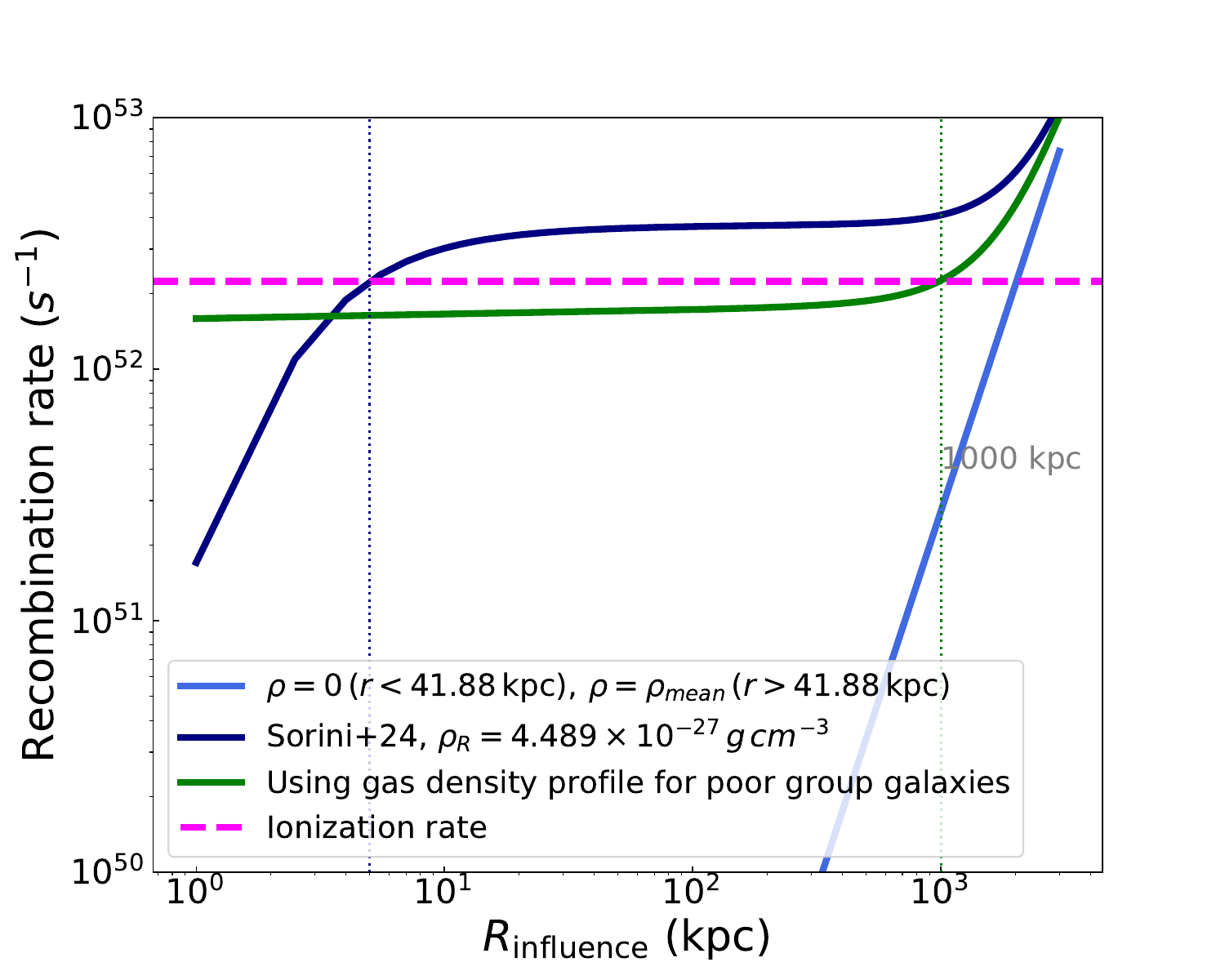}
     \caption{Total recombination rate within a given radius from the nucleus of Cen A at $z=5$. The recombination rate rises rapidly at small radii due to the high density of gas in this region. Beyond the virial radius, where density is much lower, the total recombination rate rises only slowly, until larger radii (corresponding to much larger volumes) are reached. The dark blue line shows results using a density profile based on the work of \cite{Sorini+2024} plus the mean background. The green line shows results using the density profile for poor group galaxies plus the mean background. The lighter blue line shows a model in which we set the density to zero inside $R_{200}$, with the mean density at larger radii. The horizontal, dashed magenta line indicates the ionizing rate of our model AGN at $z=5$.}
     \label{fig:recomb_rate}
 \end{figure}
 
 The right-hand side of equation~(\ref{eq:Rinfluence}) above yields the total recombination rate within a given radius of influence. Figure \ref{fig:recomb_rate} shows this recombination as a function of $R_\mathrm{influence}$. The dark blue curve shows results using the density profile given in equation~(\ref{eq:sorini}). The recombination rate increases rapidly at small radii due to the high densities in this region. Beyond the virial radius the total recombination rate grows only slowly, due to the lower densities, until sufficiently large radii (and correspondingly large volumes) are reached. The horizontal dashed line indicates the ionizing rate of our Cen A AGN at $z=5$--- where this intersects with the dark blue line indicates the radius of influence of the AGN. Note that the radius of influence is small for the recombination rate calculated with \cite{Sorini+2024}--- just a few kpc. However, it is apparent from Figure~\ref{fig:recomb_rate} that small increases in the ionizing luminosity, or decreases in the density of the gas around Cen A, could lead to a substantially larger radius of influence. For comparison, the lighter blue line in Figure~\ref{fig:recomb_rate} shows a model in which we set $n(r)=0$ inside the virial radius, and $n(r) = \rho_\mathrm{mean} X/m_\mathrm{H}$ at larger radii. Such a model may be appropriate if the gas within Cen A's virial radius has been shock heated to sufficiently high temperature to become collisionally ionized (and, therefore, does not contribute to absorbing the ionizing luminosity produced by the AGN). In this case, a radius of influence of around 2~Mpc is found. Given these results, $R_{\mathrm{influence}}$ calculated with \cite{Sorini+2024} density profile is too small to produce results that explain the Cen A luminosity function. The density profile using poor group galaxies results in the smallest radius of influence ($R_\mathrm{influence}\sim 1000\,\mathrm{kpc}$) that produces an adequately large quenching effect, and this value is consistent with gas that is substantially but not entirely shock heated. Therefore, we adopt $R_\mathrm{influence}\sim 1000\,\mathrm{kpc}$ as a typical value attained when the AGN is sufficiently luminous such that the ionization front breaks out of the halo at $z=5$. This value is closer to the noted extent of Cen A’s radio lobes in the literature  as Cen A’s radio lobes can extend as far as 600 kpc\citep{Yang+2012}.
\\

Next, we show luminosity functions corresponding to our assumed density profiles and their resulting radius of influence calculated. Two models are shown. First the model assuming $\mathrm{R_{influence}=5\,kpc}$: this model quenches satellites only within 5 kpc (only 2 satellites are quenched in this case, and they are not within 150 kpc z=0). Therefore, this is equivalent to our default Cen A model without AGN reionization effects. The remainder of the satellites are not influenced by the AGN at ; all galaxies are traced to z=0. The blue curve shows the luminosity function for these traced satellites within 150 kpc at z=0. Second, we show the model assuming $\mathrm{R_{influence}=1000\,kpc}$ using the poor group density profile, such that all satellites within $1,000\,\mathrm{kpc}$ at  are suppressed due to the AGN. This model is equivalent to AGN reionization model presented in this work. All galaxies are traced to z=0 and their luminosity function within 150 kpc at z=0 is shown in green. Note that this green curve is closer to the observed luminosity function.

\begin{figure}[h!]
     \centering
     \includegraphics[width=\linewidth]{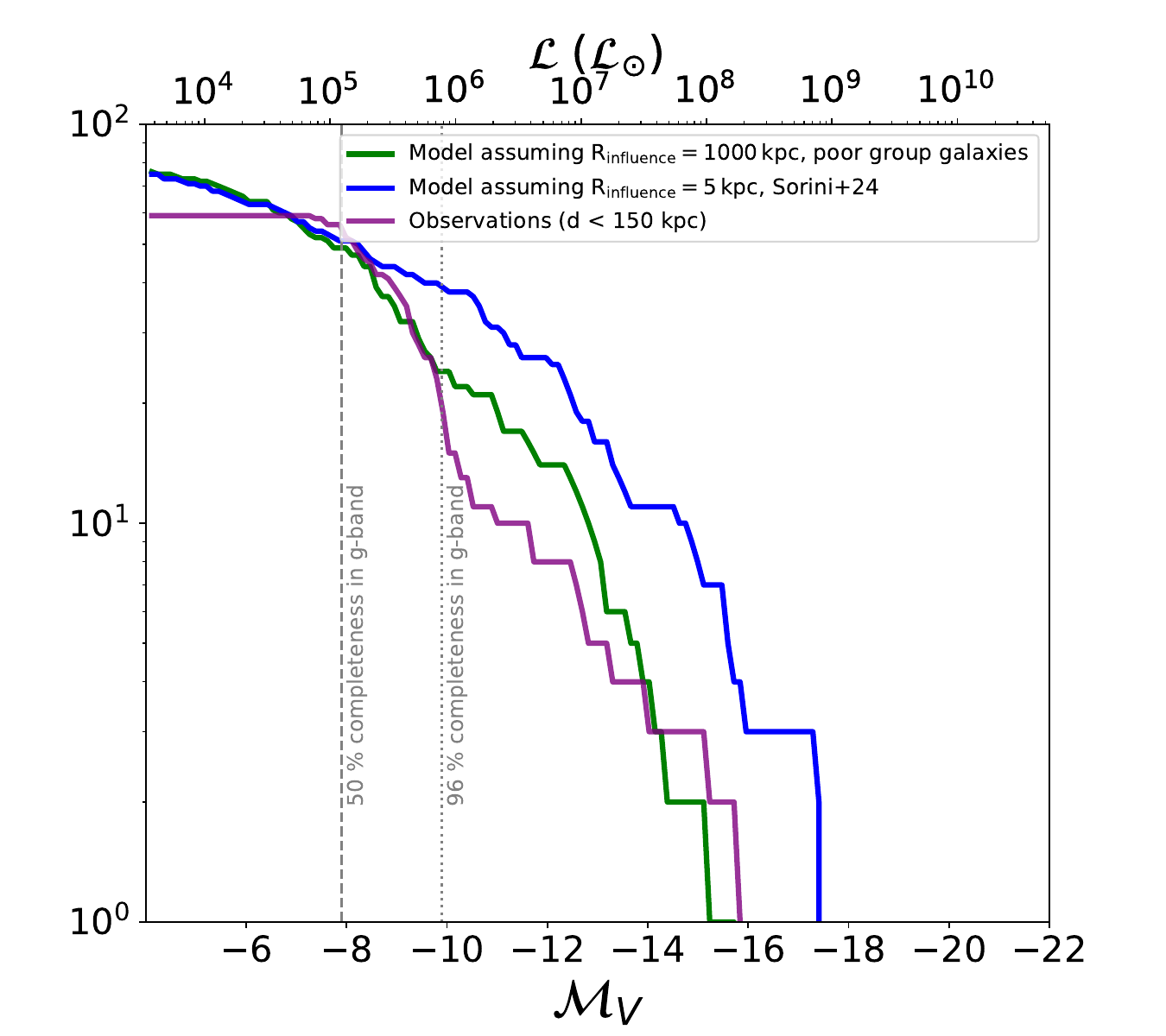}
     \caption{Luminosity function for Cen A satellite progenitors within radius of influence calculated at z=5 traced to z=0 and within 150 kpc at z=0. Blue curve shows a model assuming the density profile of \cite{Sorini+2024} and $\mathrm{R_{influence}=5\,kpc}$ and the green curve shows a model assuming $\mathrm{R_{influence}=1000\,kpc}$. LF within 150 kpc is shown in purple.}
     \label{fig:LF_densityProfiles}
 \end{figure}
 
 Note that the poor group galaxy density profile was calibrated using observations of present-day groups with halo masses substantially larger than that of the Cen A progenitor at z=5. Applying this profile to our system, therefore, constitutes an extrapolation in both redshift and halo mass that is not constrained by observations at this epoch and mass scale. We adopt it nonetheless because it is the only available density profile motivated by the specific environment of Cen A, a poor, sparse group \cite{Karachentsev+2007}, and because the same functional form has been used in prior studies of the present-day intragroup medium surrounding Cen A's giant lobes \citep{OSullivan+2013,Stawarz+2013}. However, given the sensitivity of radius of influence to the assumed density profile in \ref{fig:recomb_rate}, the resulting radius of influence should be interpreted as an order-of-magnitude estimate rather than a precise prediction.
 
 These results indicate that an AGN similar to that of Cen A has sufficient energetic output to heat the surrounding IGM\footnote{At these high temperatures, we estimate the recombination timescale in the IGM to be about 57.8~Gyrs, and the cooling timescale due to atomic processes to be around 2.472~Gyrs. Therefore, we expect the IGM surrounding Cen A satellites to remain hot and prevent accretion into satellites for a substantial time.}. At $z=5$, this feedback mechanism can influence neighboring halos at distances up to 1000~kpc, potentially regulating star formation in surrounding dwarf galaxies.

% \begin{figure}
% \centering
%     \includegraphics[width=.5\linewidth]{277kpc@z5_distributions_z0.png}
%     \caption{The radial distribution of galaxies found within z=4, 3, 2, 1 is traced to z=0. Here, we select Cen A satellites within the radius of influence of Cen A's AGN and look at their radial distribution at $z=0$ (shown in green histogram).}
% \end{figure}

%We find a better fit for the observed luminosity function for hot IGM temperature,  $V_\mathrm{quench}=80\,\mathrm{km/s}$ ($T_\mathrm{vir}=2.29\times10^5\,\mathrm{K}$) compared to our previous model from \cite{Weerasooriya+2024}. This suggests that the IGM around dwarf satellites of Cen A may have been very hot, potentially preventing accretion onto the dwarfs. 

 \begin{figure}[!ht]
    \centering
\includegraphics[width=\linewidth]{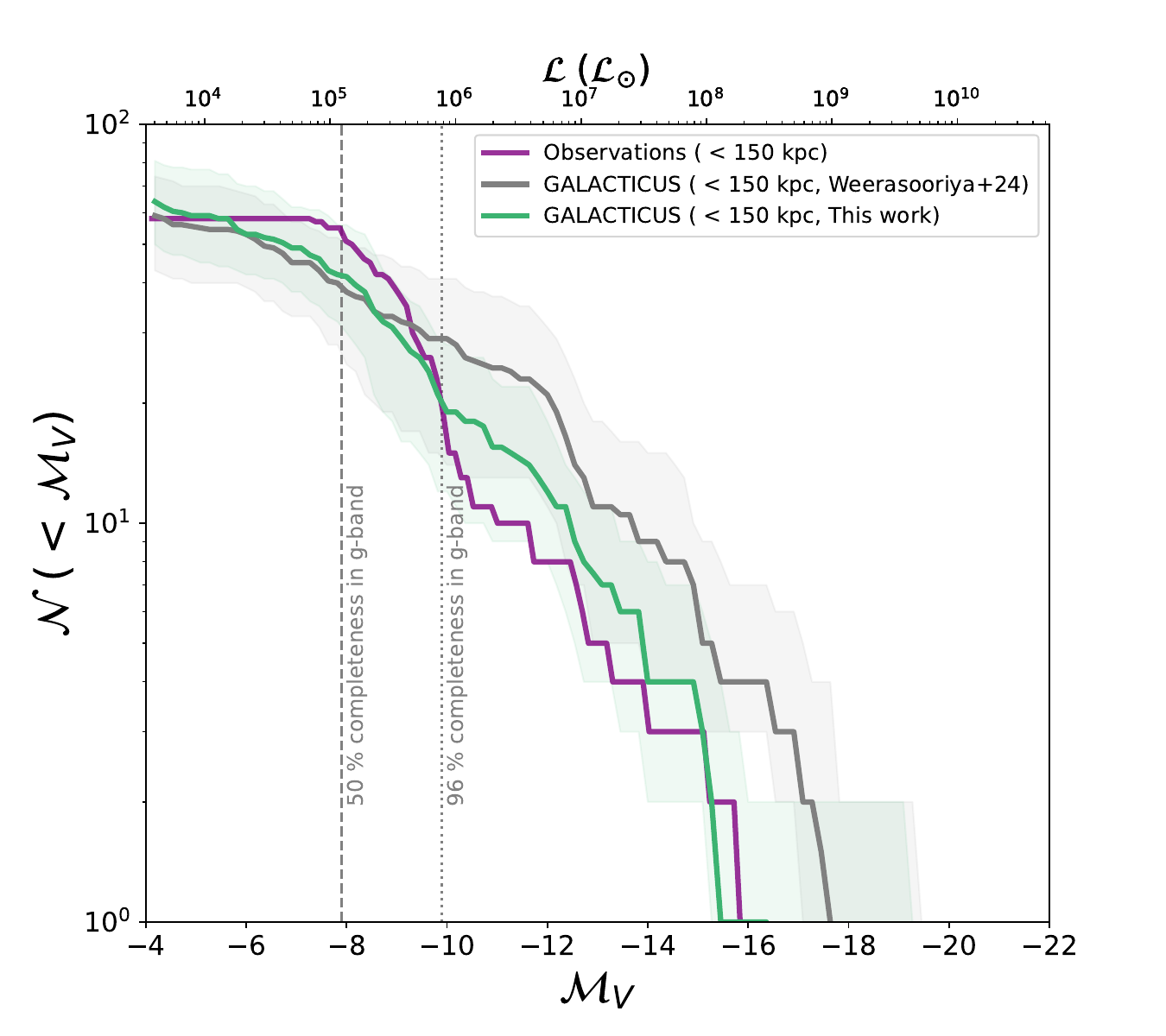}
    \caption{Our final luminosity functions for the Cen A satellites within a projected radius of 150~kpc when quenched after $z=5$ in comparison to observed luminosity functions (purple line). In comparison, the previous work of \cite{Weerasooriya+2024} (grey curve) overpredicts the number of luminous satellites. The luminosity function from this work, including the influence of Cen A's AGN, is shown by the green line and matches well with the observed luminosity function within 150 kpc}
    \label{fig:lum_Mv_final}
\end{figure}
 
 Figure \ref{fig:lum_Mv_final} shows the final luminosity functions we obtain from our \textsc{Galacticus} models for satellites within 150 kpc (in sea green) compared to the observed LF. This model provides a better match to observations than the previous model from \cite{Weerasooriya+2024}, achieved by adjusting the hot IGM temperatures ($V_\mathrm{quench}=80\,\text{km/s}$ and z=5).

\section{Discussion \& Summary} \label{sec:summary}

In this paper, we attempt to reproduce the cumulative luminosity function of Cen A satellites within $150\,\mathrm{kpc}$, a system that is notable for having no satellites brighter than $\mathrm{M_\mathrm{V}}\leq-15.7$. This strong cutoff motivates us to explore physical mechanisms that could suppress the formation or survival of bright satellites. We find that suppressing gas accretion onto satellite galaxies at a very early redshift, due to elevated local IGM temperatures resulting from AGN activity, can naturally explain this feature. This suggests that dwarf galaxies around AGN host systems may retain signatures of past AGN heating, making their luminosity functions a potential diagnostic of AGN influence.

Observational evidence supports the idea that Cen A’s AGN may affect nearby satellites. \cite{Johnson+2015} show that one of the brightest satellites within $150,\mathrm{kpc}$, ESO 324-G024 ($M_\mathrm{V} \sim -15.6$), lies behind the northern radio lobe of Cen A. They find that the HI tails of Cen A dwarf galaxy ESO 324-G024  is most likely formed through ram pressure stripping when passing through the hot ionized gas of radio lobes in their study. The authors arrive at this conclusion through a Faraday depolarization/rotation measure analysis of synchrotron radiation from lobes, which constrain the dwarf's position relative to the radio lobe, combined with explicit calculations ruling out other mechanisms like tidal stripping. This ram pressure stripping scenario is lobe specific, and the ambient CGM alone could not produce the HI tail feature of ESO 324-G024. They find that the minimum IGM density required to create the HI tail through ram pressure stripping is $10^{-3}\,\mathrm{cm^{-3}}$ if the lobe is at rest and $10^{-4}\,\mathrm{cm^{-3}}$ if the lobe is moving at $v_{lobe}=197\,\mathrm{km/s}$. This is a direct example of a satellite being influenced by AGN-driven structures, strengthening the motivation for our approach.

Recent studies have begun exploring the effects of AGN on dwarf galaxies. \cite{Visser-Zadvornyi+2025} argue that AGN heating can suppress gas accretion, quenching satellites, while \cite{Dashyan+2019} show that AGN feedback can increase the temperature and relative velocity of the IGM, also promoting quenching. However, neither study connects AGN feedback to observable changes in the satellite luminosity function, a key gap our work begins to address.

Although we currently have only one clearly affected Cen A analog, our method offers a way to probe the influence of AGN using satellite luminosity functions. In light of upcoming observations with the Roman Space Telescope and Rubin observatories, this quenching model can be applied to other host galaxies with AGN. This will allow us to test this method and also apply it as a tool to detect the presence of AGN activity using this work. Next-generation wide-field surveys (e.g., Roman, Rubin) will enable tests of this quenching model across multiple AGN host systems, increasing statistical power for this technique.\\

Finally, we acknowledge that we do not test the effects of a major merger on the luminosity function of Cen A. Given Cen A's recent major merger 2 Gyrs ago \citep{Wang+2020}, testing this possibility is necessary. Our Cen A analog has undergone a major merger at 1.65 Gyrs. A detailed study of merger history on Cen A’s luminosity function would require a large statistical sample of EPS merger trees that are robustly constrained. This work is currently in preparation by Wanichwecharungruang, Nadler \& Benson, in prep.

\noindent {Our conclusions are as follows:
\begin{itemize}

    \item We find a deficit in luminous satellites (brighter than $\mathrm{M_\mathrm{V}} < -15.8$) within 150 kpc of Cen A.
    
    \item We used the semi-analytic framework \textsc{Galacticus} (based on \cite{Weerasooriya+2024} to explore different quenching pathways for bright Cen A satellites within 150 kpc.

    \item Enhanced tidal stripping of the ISM reduces overall satellite abundances, but does not change the luminosity function slope.
    \item Enhanced tidal destruction of halos flattens the luminosity function at the faint end, effectively reducing the slope.
    \item Earlier reionization flattens the luminosity function at the fainter end.
    \item Reionization-driven suppression/hotter IGM can reproduce the correct shape of the luminosity function.
    
    \item Suppressing gas accretion due to AGN, where the IGM is heated up to $\sim2.3 \times 10^5$ K at z = 5, can reproduce the observed luminosity function of the Cen A system. This suggests that the AGN activity from the main halo could plausibly be responsible for heating the surrounding IGM.

\end{itemize}

In summary, our work suggests that dwarf satellites around galaxies with AGN can reveal imprints of past AGN activity. The follow-up study will explore star formation histories of dwarfs around AGN hosts. Although further observations are needed, considering possible merger-induced quenching of Cen A satellites provides a compelling case study, and forthcoming deep surveys will substantially expand our ability to test this method across a broader host population.\\

\noindent We thank the referee for their thoughtful and constructive feedback provided during the review process. The authors acknowledge the OBS HPC at Carnegie Science Observatories, and the Resnick High Performance Computing Center, a facility supported by the Resnick Sustainability Institute at the California Institute of Technology.\\

\noindent \textit{Software:} \textsc{Galacticus} \citep{Benson2012}, \texttt{CONSISTENT\_TREES} \citep{consistent_trees}, \texttt{AMIGA} \citep{AHF}, \texttt{JUPYTER} \citep{Jupyter}, \texttt{MATHEMATICA} \citep{Mathematica}, \texttt{NUMPY} \citep{numpy}, \texttt{SCIPY} \citep{scipy}, and Matplotlib \citep{matplotlib}.

\bibliography{sample631,ref2.bib,ref}{}
\bibliographystyle{aasjournal}

%% This command is needed to show the entire author+affiliation list when
%% the collaboration and author truncation commands are used.  It has to
%% go at the end of the manuscript.
%\allauthors

%% Include this line if you are using the \added, \replaced, \deleted
%% commands to see a summary list of all changes at the end of the article.
%\listofchanges
\appendix

\section{Can these AGN-influenced galaxies still remain within the inner regions of Cen A at z=0}

\begin{figure*}[t]
\subfloat[]{
\includegraphics[width=\linewidth]{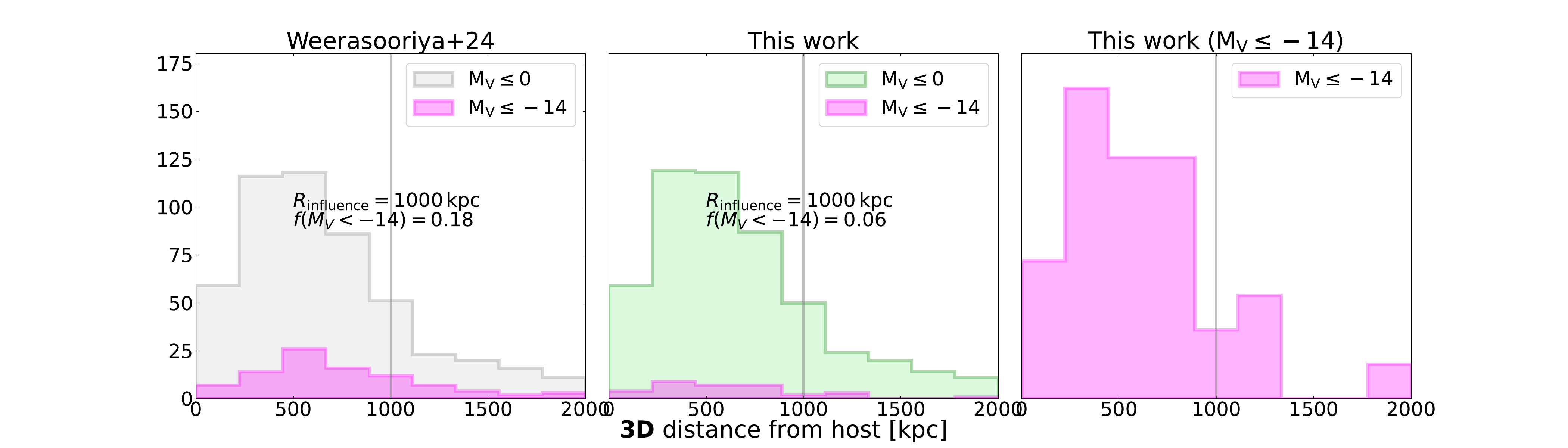} 
}\hfill
\subfloat[]{
\includegraphics[width=\linewidth]{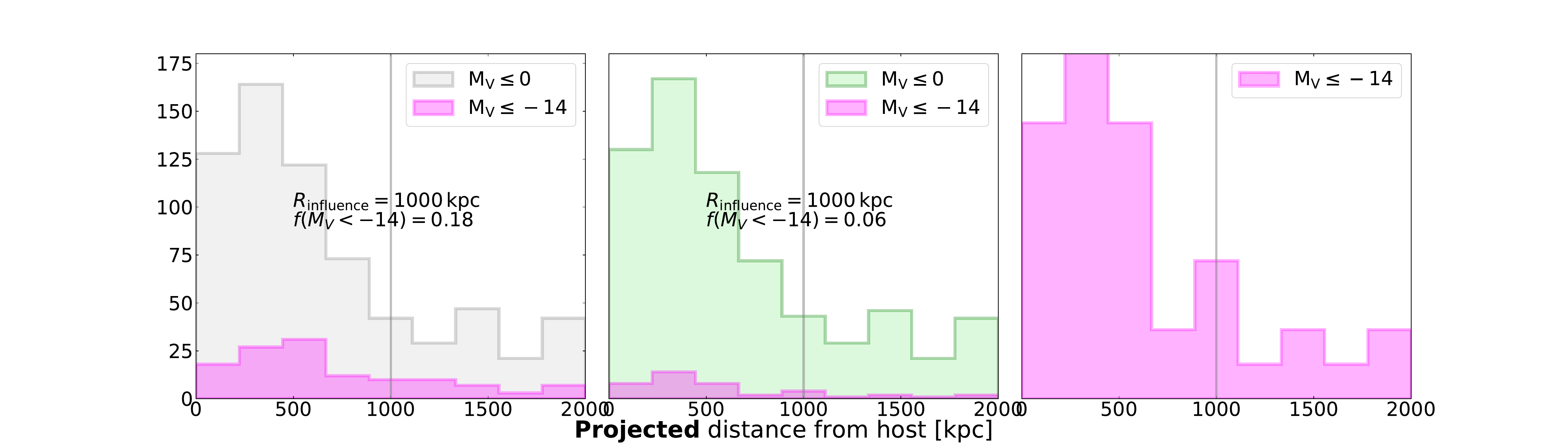} 
}
     \caption{The radial distribution of dwarf galaxies found within $R_\mathrm{influence}=1000\,\mathrm{kpc}$ at $z=5$, traced to $z=0$. Top and bottom rows shows galaxies within a 3D distance, and a 2D projected distance from the host respectively. Results from the model of \cite{Weerasooriya+2024} are shown in the left column, while results from this work including quenching by the Cen A AGN are shown in the middle column. Dwarfs brighter than $M_\mathrm{V} < 0$ are shown in grey and green, respectively in the two columns, with galaxies brighter than $M_\mathrm{V} < -14$ shown in pink. The vertical grey lines indicate the approximate radius of influence of the AGN. In the right column we zoom in to just those galaxies brighter than  $M_\mathrm{V}\sim-14$ in the model from this work.}
     \label{fig:distributions}
\end{figure*}

Finally, we examine where dwarfs forming within the radius of influence end up at later epochs. In Figure \ref{fig:distributions}, we track dwarfs that were within the radius of influence at $z=5$, and, for those which survive to the present time (i.e., have not been destroyed or merged with the central galaxy) examine their distribution of 3D and 2D projected (top and bottom rows, respectively) distances from Cen A at $z=0$. The results are shown for the model of \cite{Weerasooriya+2024} in the left column, and for this work (including the quenching effects of the AGN) in the middle column. In each case, the pink histograms show results for galaxies brighter than $M_\mathrm{V}=-14$ at $z=0$, while the lighter histograms (grey in the left column, green in the middle column) show results for galaxies brighter than $M_\mathrm{V}=0$. The right column repeats the results from this work, but zooms in to the population of bright satellites.  We note that bright satellites ($M_\mathrm{V} < -14$ make up 18\% of the total ($M_\mathrm{V}<0$) population in the model of \cite{Weerasooriya+2024}, but only 6\% in our model including the effects of the Cen A AGN. Figure~\ref{fig:distributions} clearly shows that bright galaxies within the radius of influence at $z=5$ are primarily found within a similar range of radii at $z=0$, biased toward smaller radii (i.e. the distribution in the right panel of Figure~\ref{fig:distributions} peaks at 300--400~kpc). Furthermore, the number of bright galaxies at all radii is seen to be suppressed (relative to the model of \citealt{Weerasooriya+2024}) by the quenching effects of the AGN.

\section{Impact of host mass}\label{sec:massDep}
\begin{figure}[h!]
    \centering
    \includegraphics[width=.6\linewidth]{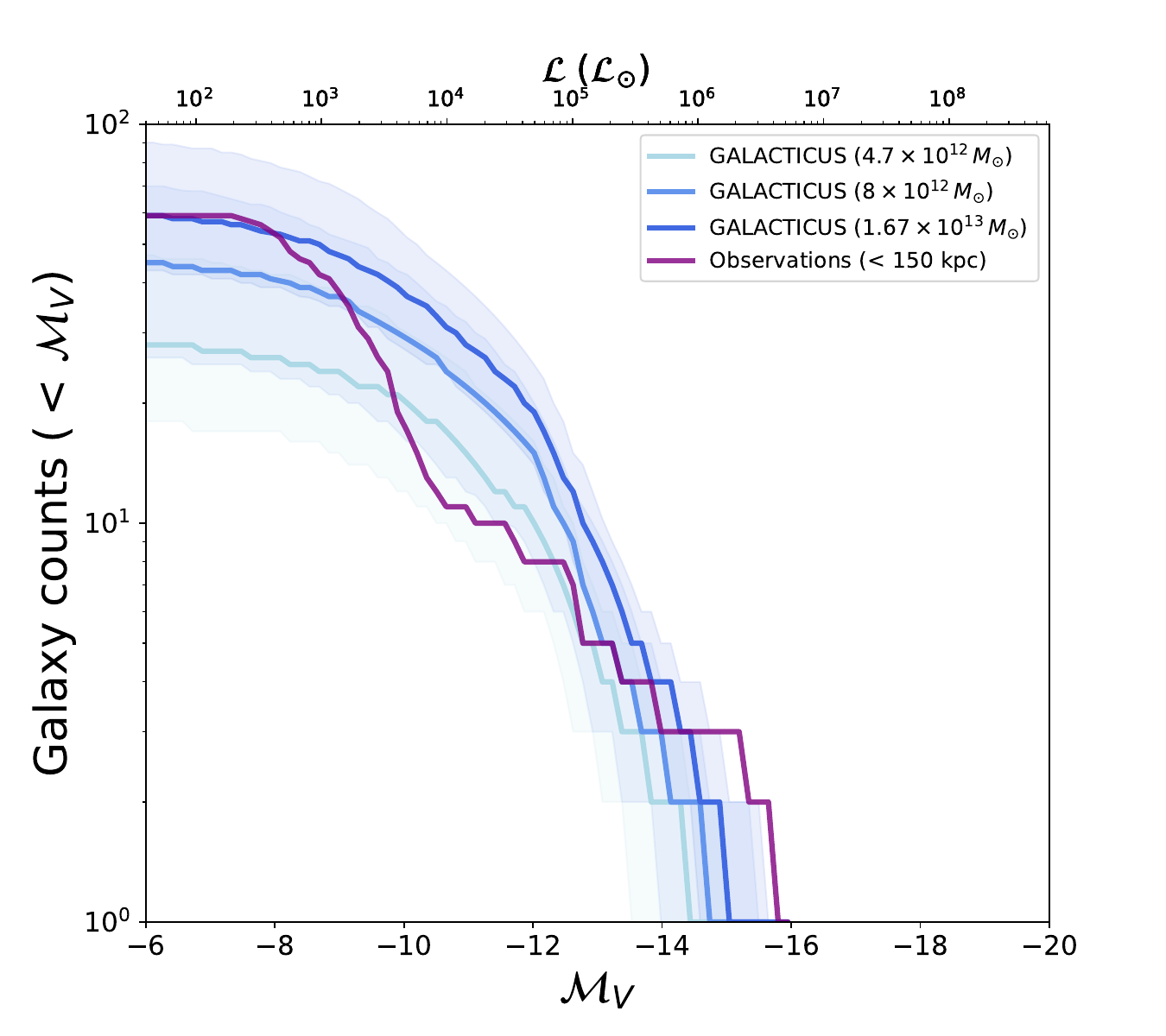}
    \caption{Cumulative luminosity functions of Cen A satellites within 150~kpc projected radius for Cen A analogs with $4.7\times 10^{12}\,\mathrm{M}_{\odot}$ (light blue), $8\times 10^{12}\,\mathrm{M}_{\odot}$ (cornflower blue), and $1.67\times 10^{13}\,\mathrm{M}_{\odot}$ (royal blue). Observations are shown by the purple line. Solid lines show the median (over 100 merger tree realizations and all rotations of the system around the $x$-axis) number of galaxies brighter than a given magnitude, and the shaded regions show the 25th and 75th percentiles of the distribution over realizations and rotations.}
    \label{fig:masses}
\end{figure}

The mass of Cen A remains poorly constrained with a lower limit of $4.7\times 10^{12}\,\mathrm{M}_{\odot}$ \citep{Pearson+2022} and an upper limit $1.8\times 10^{13}\,\mathrm{M}_{\odot}$\citep{van2000}. We have only a single N-body simulation of a Cen A analog, which has a mass of $1.67\times10^{13}\,\mathrm{\mathrm{M}_{\odot}}$. Therefore, to test the impact of different Cen A analog masses on the luminosity function of satellites we generate a total of 300 merger trees with halo masses $4.7\times10^{12}\,\mathrm{\mathrm{M}_{\odot}}$, $8\times10^{12}\,\mathrm{\mathrm{M}_{\odot}}$, and $1.67\times10^{13}\,\mathrm{\mathrm{M}_{\odot}}$ using the merger tree building algorithm of \cite{Cole2000} based on the Extended Press-Schechter (EPS) formalism with the modifier function of \cite{Parkinson+2008}.

Figure \ref{fig:masses} shows the cumulative luminosity functions for the $4.7\times 10^{12}\,\mathrm{M}_{\odot}$, $8\times 10^{12}\,\mathrm{M}_{\odot}$, and $1.67\times 10^{13}\,\mathrm{M}_{\odot}$ Cen A analogs in shades of blue with the solid lines showing the median over all 100 merger trees of each mass with shaded regions indicating the range of variation over realizations and over rotations of each realization around the $x$-axis. It can be seen that the $1.67\times 10^{13}\,\mathrm{M}_{\odot}$ does not produce as many of the brightest satellites as our N-body merger tree, even though the total mass of Cen A is the same. This suggests that either the merger history of our N-body simulation is unusual (although this seems unlikely as its luminosity function lies well outside of the shaded region in Figure~\ref{fig:masses}, suggesting that it is not simply an outlier in the distribution of luminosity functions predicted by these extended Press-Schechter trees), or that these extended Press-Schechter merger trees do not accurately capture the formation of the most massive subhalos in these Cen A analog halos. However, here we use these extended Press-Schechter merger tree results simply to quantify the relative change in the luminosity function as the mass of Cen A is changed.

It can be seen that the overall normalization of the cumulative luminosity function depends on the mass of Cen A, with more massive Cen A's having more satellites overall. The median number of brighter satellites ($M_\mathrm{V}\leq-14$) increases from 9 to 15 as the mass increases from $4.7\times 10^{12}\,\mathrm{M}_{\odot}$ to $8\times 10^{12}\,\mathrm{M}_{\odot}$ and from 15 to 23 satellites when the mass of the host increases from $8\times 10^{12}\,\mathrm{M}_{\odot}$, $1.67\times 10^{13}\,\mathrm{M}_{\odot}$. Assuming the same relative change in the luminosity function obtained from N-body merger trees if the mass of the Cen A analog were similarly reduced, it is clear that a mass toward the lower end of the allowed range for Cen A could result in a match to the observed number of galaxies brighter than $M_\mathrm{V}\leq-14$. However, as can be seen from Figure~\ref{fig:masses}, this would also reduce the number of fainter satellites, meaning that such models would then fail to match the observations at faint magnitudes. As such, simply reducing the mass of the Cen A system is not a viable solution to the observed dearth of bright satellites.

\section{Comparison of models and observations} \label{appendix2}
\subsection{Modeled vs. observed satellites}
\begin{figure}
    \centering
    \includegraphics[width=.6\linewidth]{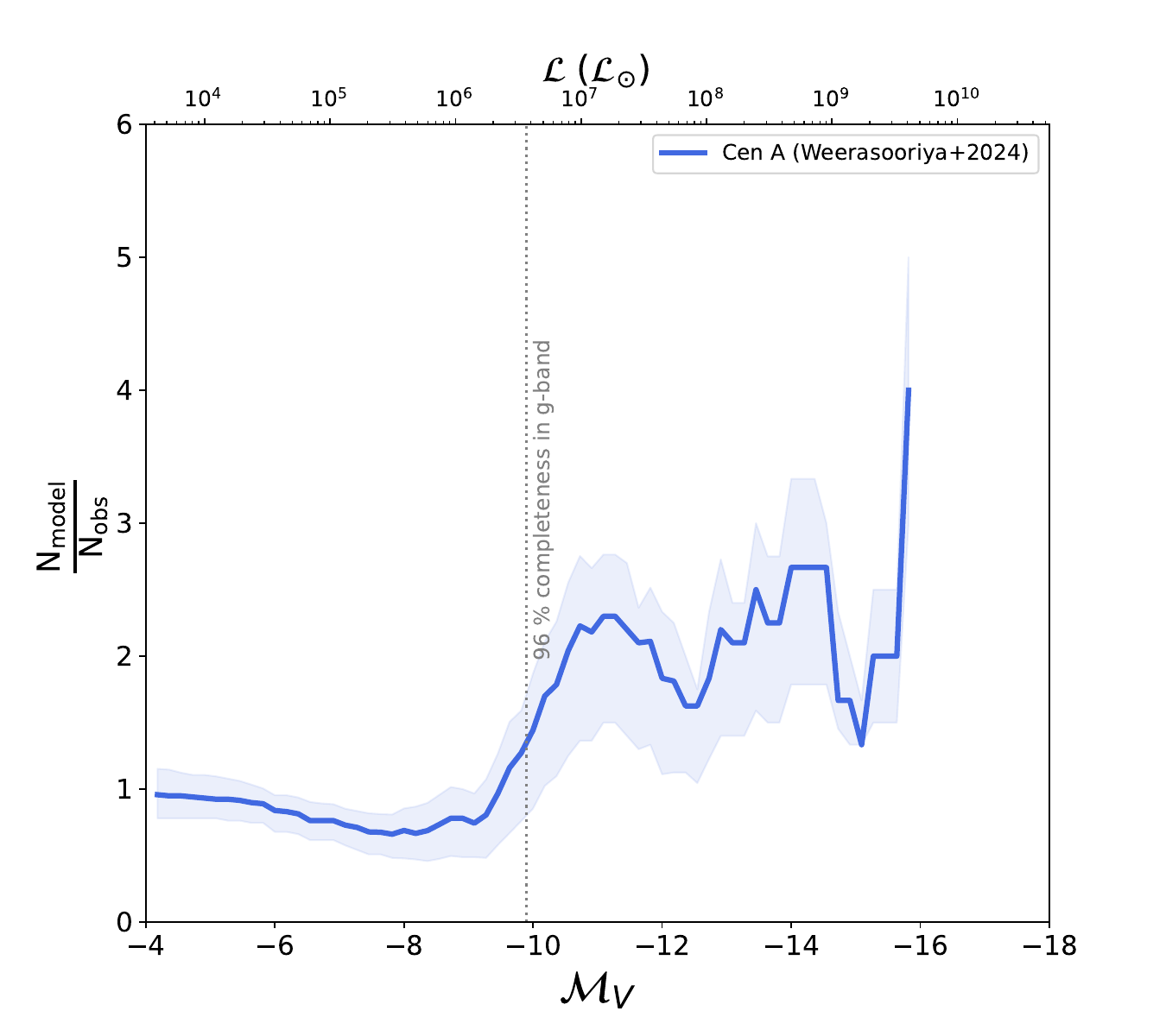}
    \caption{Comparison of cumulative number of modeled satellites to observed number of satellites at V band magnitude. Solid blue curve shows the median of $\mathrm{N_{models}/N_{observed}}$ and the shaded region shows 1 sigma envelope.}
    \label{fig:comparisonLF}
\end{figure}
To provide another perspective of the significance of the difference in the number of observed vs. modeled satellites, we compare the cumulative number of modeled satellites to the observed number of satellites within 150 kpc at each absolute V band magnitude. Figure \ref{fig:comparisonLF} shows the ratio of the number of galaxies observed in our fiducial model \citep{Weerasooriya+2024} to the number of Cen A satellites observed. The solid curve shows the median value of this ratio while the shaded region shows the $1 \sigma$ envelope over all rotations of the simulation. We find that our models from \cite{Weerasooriya+2024} over-predict the number of satellites at $M_\mathrm{V}\sim-15.8$ by a factor of $4\pm1$ and by a factor of $1.8-3.3$ at $M_\mathrm{V}\sim-14$. 

\subsection{Statistics of the number of bright satellites}\label{sec:stats}

\begin{figure}[ht!]
    \centering
    \includegraphics[width=.6\linewidth]{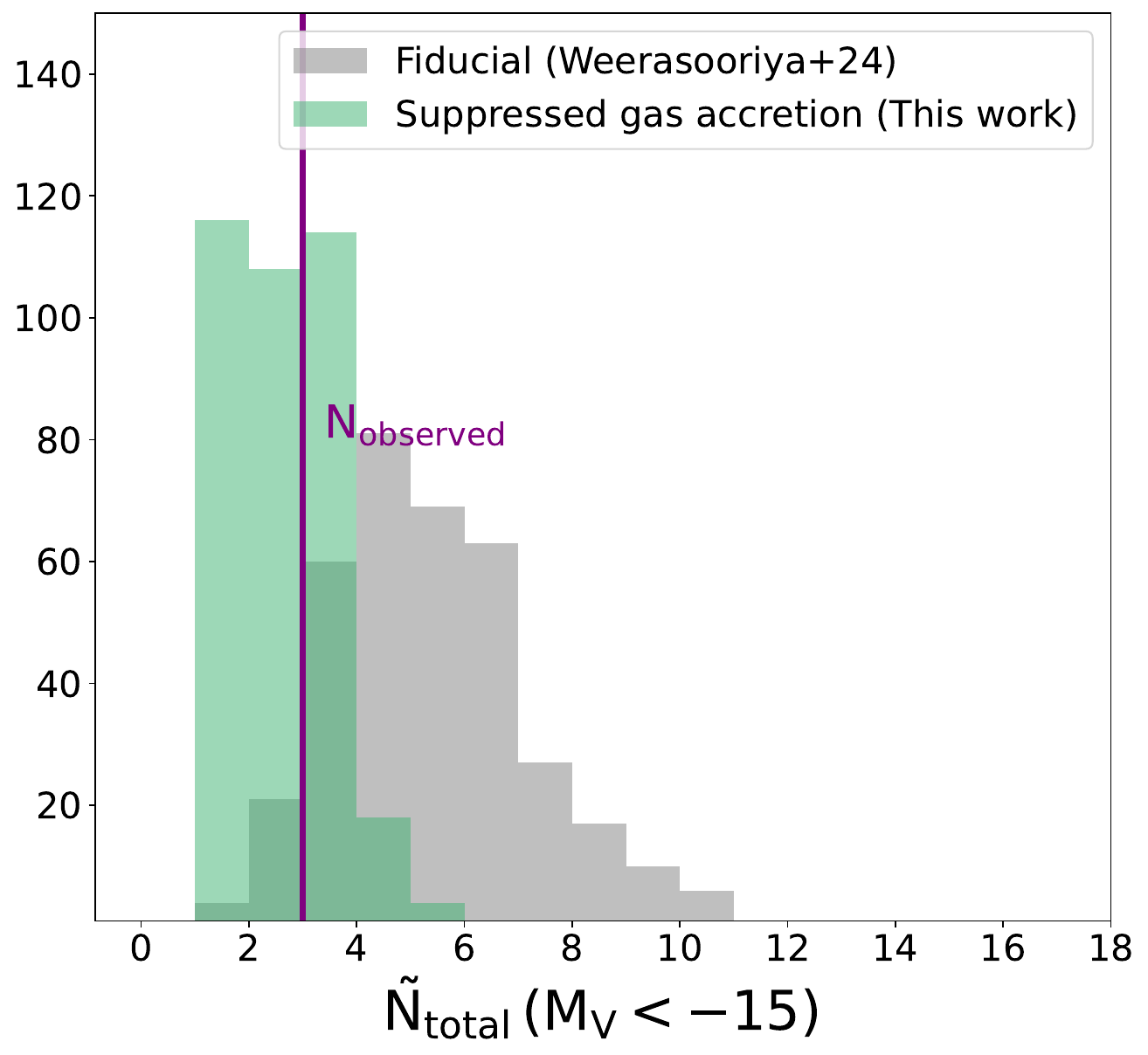}
    \caption{The distribution of total number of bright satellites, $\mathrm{N_{total,random}}$, having $\mathrm{M_\mathrm{V}}<-15$ for the fiducial model of \citeauthor{Weerasooriya+2024}~(\citeyear{Weerasooriya+2024}; gray) and the final model (including the quenching effects of Cen A's AGN) of this work (green). The vertical, purple line indicates the observed number of such galaxies, $N_\mathrm{observed}=3$.}
    \label{fig:probability}
\end{figure}

The N-body simulation used in this paper provides only one realization of the Cen A system. If many realizations were available, each would show a different number of satellite galaxies, with variations arising due to both differences in formation history and from random variations arising from the specific spatial distribution of satellites combined with the direction from which the system is viewed. In this work, we take the latter effect into account by rotating the N-body system around an axis to simulate different viewing angles. However, since we are concerned with luminosity functions of galaxies that lie within some \emph{projected} radius, $R$ of Cen A, such rotations will never change the number of galaxies contributed by the region within a 3D radius $r=R$. To account for this issue, here we resample the number of satellites within this $r=R$ sphere, assuming Poisson statistics as we rotate the satellite system. While this does not fully account for variations related to differences in merger histories, \cite{2010MNRAS.406..896B} showed that the distribution of the number of subhalos in a halo follows a negative binomial distribution, which, for small numbers of subhalos, is well-approximated by a Poisson distribution.

We therefore proceed as follows. We first count the number of satellites brighter than $M_\mathrm{V}<-15$ within a 3D distance of 150~kpc from Cen A in our model, defining this as $N_\mathrm{inner}$. For each rotated configuration, we draw a random number of inner satellites from a Poisson distribution with mean $\lambda=N_\mathrm{inner}$, yielding $N_\mathrm{inner, i}$. Next, we count the number of galaxies with $M_\mathrm{V}<-15$ in a 2D projected radius of $150\,\mathrm{kpc}$ for the current rotation, calling this $N_\mathrm{total, i}$. The number of galaxies in the projected radius of 150~kpc, but \emph{not} within the 3D radius of 150~kpc is then $N_\mathrm{outer, i} = N_\mathrm{total, i} - N_\mathrm{inner}$. Finally, we construct the resampled, total number of galaxies as $\tilde{N}_\mathrm{total, i} = N_\mathrm{outer, i} + N_\mathrm{inner, i} =  N_\mathrm{total, i} - N_\mathrm{inner} + N_\mathrm{inner, i}$. We repeat this process for each rotation of the simulation box, incremented in one-degree steps.

Figure~\ref{fig:probability} shows the resulting distribution of $\tilde{N}_\mathrm{total}$ for both the fiducial model from \citet{Weerasooriya+2024} and our AGN-suppressed gas accretion model. The vertical line indicates the observed number of galaxies with $M_\mathrm{V}<-15$ in a 2D projected radius of $150\,\mathrm{kpc}$, $N_\mathrm{observed}=3$.  While $N_\mathrm{observed}=3$ is consistent with the low-$\tilde{N}_\mathrm{total}$ tail of the distribution for our fiducial model, it is clearly quite unlikely. In contrast, $\tilde{N}_\mathrm{total}$ is highly probable in our AGN suppressed gas accretion model (which predicts that the vast majority of Cen A halos will have 1--3 such galaxies, with a small minority having up to 5). 

This analysis suggests that, if many independent realizations of the Cen A system were available, configurations with $\tilde{N}_\mathrm{total, i}$ would occur much more frequently in scenarios where external AGN activity suppresses gas accretion onto galaxies.

\subsection{Black hole mass--velocity dispersion relation}

\textsc{Galacticus} has been shown to produce reasonable black hole masses and velocity dispersions for Cen A mass halos. Figure \ref{fig:Msigma} shows a comparison of \textsc{Galacticus} models for the galaxy population in halos with virial masses in the range $10^{10}$--$10^{13}\,\mathrm{\mathrm{M}_{\odot}}$ \citep[blue line with error bars indicating the model scatter; ][in prep]{Vazquez+2025}. The observational sample of \cite{McConnel&Ma2013} is shown by grey points with the best fit relation of \cite{Kormendy&Ho2013} shown as the grey line, while the observed data for Cen A is shown by the purple circle. \textsc{Galacticus} matches the trend in black hole mass over the range of velocity dispersions for which the model was evaluated, although it is offset to masses approximately a factor of 2 too high.

\begin{figure}[ht!]
    \centering
    \includegraphics[width=0.5\linewidth]{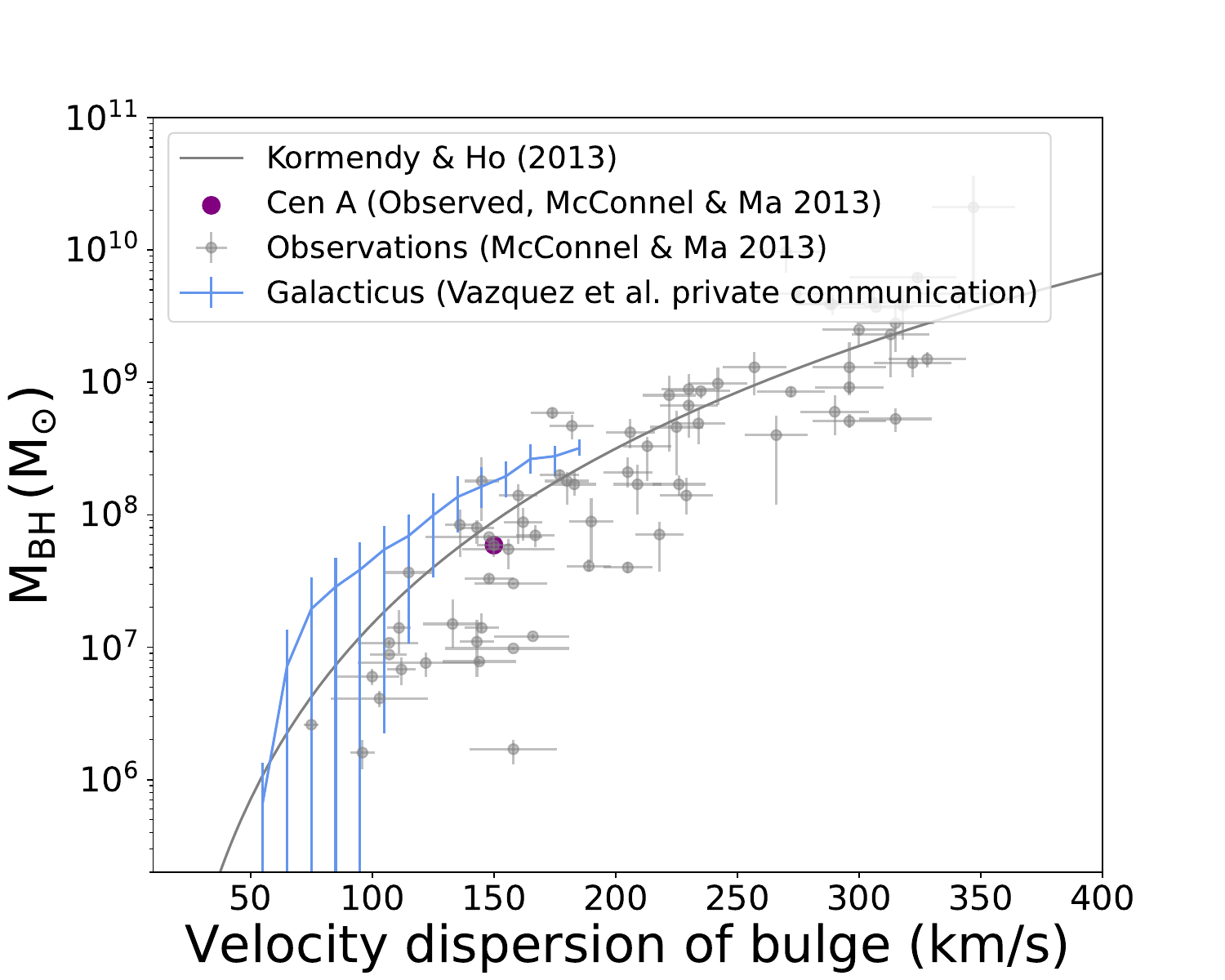}
    \caption{The black hole $M$--$\sigma$ relation for a statistical sample of galaxies generated with \textsc{Galacticus} for galaxies with $M_\mathrm{h}=10^{12}$--$3\times10^{13}\,\mathrm{M}_{\odot}$ (Vazquez et al., private communication). The median relation is shown in the solid line blue with error bars indicating the $1 \sigma$ model scatter. The observational sample of \cite{McConnel&Ma2013} is shown by grey points with the best fit relation of \cite{Kormendy&Ho2013} shown as the grey line, while the observed data for Cen A is shown by the purple circle.}
    \label{fig:Msigma}
\end{figure}

\end{document}